\documentclass[%
 aip,
 amsmath,amssymb,
 reprint,%
]{revtex4-1}

\usepackage{graphicx}  
\usepackage{dcolumn}
\usepackage{bm}

\usepackage[utf8]{inputenc}
\usepackage[T1]{fontenc}
\usepackage{mathptmx}
\usepackage{etoolbox}
\usepackage{xcolor}  
\makeatletter
\def\@email#1#2{%
 \endgroup
 \patchcmd{\titleblock@produce}
  {\frontmatter@RRAPformat}
  {\frontmatter@RRAPformat{\produce@RRAP{*#1\href{mailto:#2}{#2}}}\frontmatter@RRAPformat}
  {}{}
}%
\makeatother
\begin{document}

\preprint{AIP/123-QED}

\title{Experimental demonstrations of Josephson threshold detectors for broadband microwave photons detection}
\author{Jia-Xing He}
\affiliation{Information Quantum Technology Laboratory, 
School of Information Science and Technology, Southwest Jiaotong University, Chengdu 610031, China}
\author{Ya-Qiang Chai}
\affiliation{Information Quantum Technology Laboratory, 
School of Information Science and Technology, Southwest Jiaotong University, Chengdu 610031, China}%
\author{Peng-Hui Ouyang}
\affiliation{Information Quantum Technology Laboratory, 
School of Information Science and Technology, Southwest Jiaotong University, Chengdu 610031, China}%
\author{Hong Chang}
\affiliation{Qixing Vacuum Coating Technology Co., Ltd., Chengdu 610031, China}
\author{L. F. Wei*}
\email{lfwei@swjtu.edu.cn}
\affiliation{Information Quantum Technology Laboratory, 
School of Information Science and Technology, Southwest Jiaotong University, Chengdu 610031, China}

\date{\today}

\begin{abstract}
Current-biased Josephson junctions (CBJJs) have been demonstrated as sensitive Josephson threshold detectors (JTDs) in the infrared range. In this letter, we show this kind of detector could also be used to detect broadband microwave photons. Based on the numerical simulations of the noise-driving phase dynamics of an underdamped Josephson junction, driven by the low-frequency triangular wave current, we argue that the microwave photons flowing across the JJ can be detected by probing the voltage switched signals of the JJ. Experimentally, we designed and fabricated the relevant Al/AlOx/Al Josephson device and measured its response to microwave photons at 50~mK temperature. Experimental results indicate that the weak microwave signals could be threatened as the additional noises modify the phase dynamics of the CBJJ, which could thus be detected by the generated JTD. The detection sensitivity is characterized by using the Kumar-Caroll index to differentiate the junction switched duration distributions, with and without microwave signal input. Although the demonstrated detection sensitivity is just $-92$~dBm (corresponding to approximately 30~photon/ns) for the microwave photons at $\sim 5$GHz (which is manifestly deviated from the plasma frequency of the fabricated JJ), we argued, based on the relevant numerical simulations, that the generated JTD could be used to achieve the sensitive detection of the microwave photons at the plasma frequency of the JJ.
\end{abstract}

\maketitle

Single-photon detection technology has played more and more important roles in quantum information processing, e.g., optical quantum  computing~\cite{RevModPhys.79.135}, quantum key distribution~\cite{hadfield_single-photon_2009}, quantum precision measurements~\cite{GUO20222291,PhysRevLett.132.220801}, quantum radar~\cite{10209149}, etc.. Benefited from the low working temperature, superconducting single-photon detectors, typically inducing the superconducting nanowire single-photon detectors (SNSPD)~\cite{zhang_research_2021}, superconducting tunnel junction single-photon detectors (STJ)~\cite{friedrich_characterization_2022}, microwave kinetic inductance single-photon detector (MKID)~\cite{10.1063/5.0234649}, and superconducting transition edge single--photon detector (TES)~\cite{Daiji_FUKUDA20192018SDI0001}, etc.. have demonstrated the superior performances for the sensitive detections of photons in infrared range. As the microwave single-photon energy is significantly low~\cite{PANKRATOV2022582}, all these superconducting detectors, as well as the widely-used semiconducting single-photon detectors, however, cannot be utilized to implement the single-photon detection at microwave band, although it is desirable for the microwave quantum radar and the electromagnetic response detections of the axions~\cite{PhysRevLett.111.231801}, dark photons~\cite{PhysRevD.99.075002}, and high-frequency gravitational waves~\cite{PhysRevD.98.064028,PhysRevD.106.104003}, etc.. 

Given the detection sensitivity of the traditional microwave receiver (working at room temperature) is significantly low (due to the limit of the thermal noise)~\cite{layne2014receiver}, the implementation of microwave single-photon detection with various artificial quantum systems has been paid much attention in recent years~\cite{PhysRevLett.102.173602,inomata_single_2016}. Among them, superconducting Josephson junctions (JJs) play a particularly important role, as their plasma frequency is just in the microwave regime. In fact, there are two approaches to using the JJs to implement microwave single-photon detections. One is narrow-band detection implemented by using either the JJ-based qubit via probing the microwave single-photon induced qubit excitation or the qubit-cavity coupling via probing the qubit induced cavity-mode modification~\cite{PhysRevLett.107.217401,PhysRevB.86.174506}. The other one is the wide-band detection, typically such as the bolometer~\cite{D2NA00937D} and calorimeter~\cite{PhysRevApplied.11.054074}, implemented by probing the thermal response of the microwave single photons radiating on the JJs~\cite{PhysRevLett.93.107002,10.1063/1.4824308,REVIN2020960,PANKRATOV2022582,pankratov_towards_2022}. Physically, the so-called Josephson threshold detector (JTD)~\cite{PhysRevApplied.11.044078,ali_josephson_2024} is particularly suitable to realize such a detection; when the current through a JJ exceeds its critical current (or practically certain threshold value), the macroscopic phase of the JJ undergoes a drastic change, which results in an observable finite voltage across the junction. This implies that the current-biased JJ (CBJJ) device with the lower ``threshold'' value could be utilized to implement the sensitive detection of microwave photons~\cite{PhysRevB.111.024501}.

\begin{figure}[htbp]
  \centering
  \includegraphics[width=0.7\linewidth]{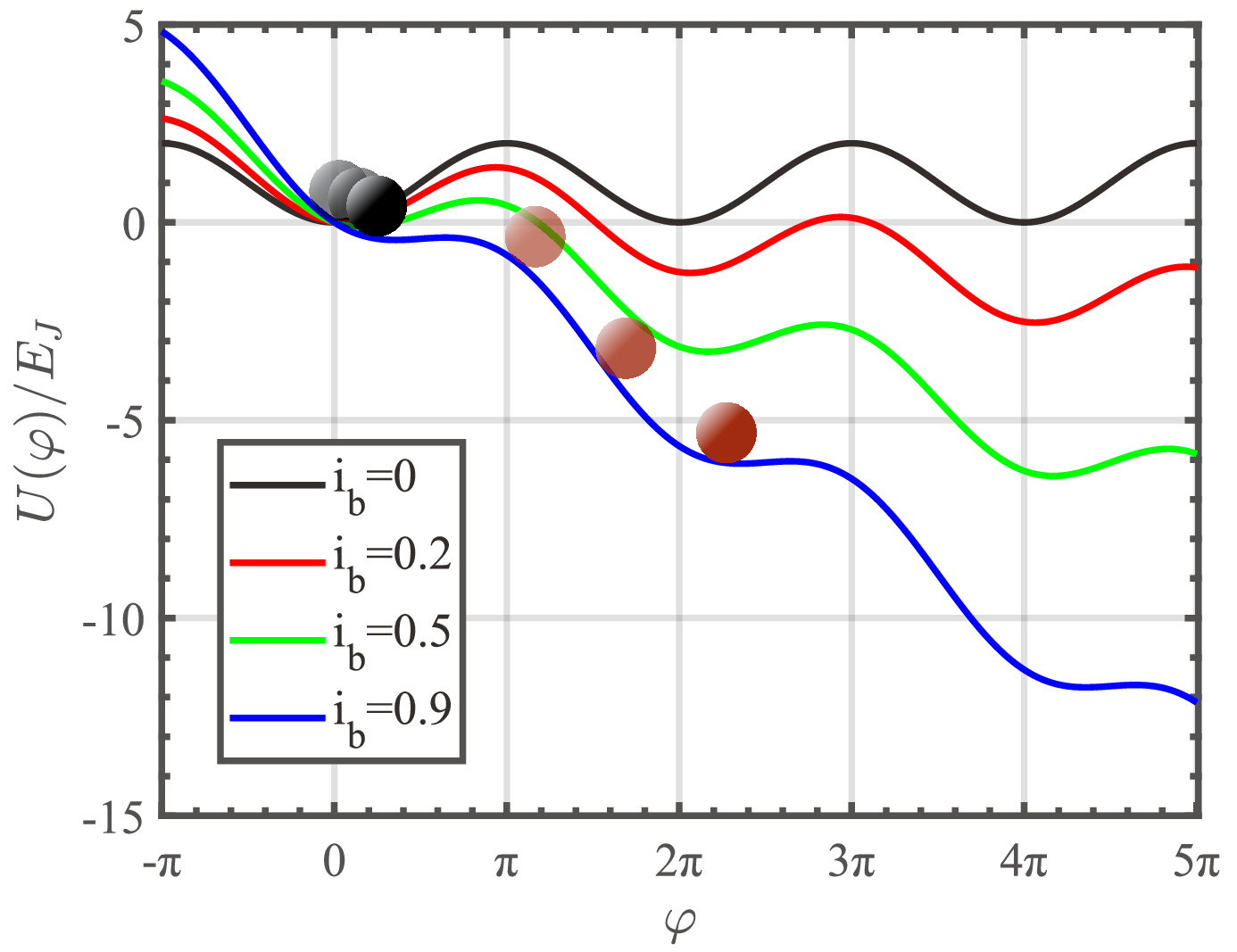}
  \caption{The ``phase particle'' dynamics description of a CBJJ device, where the black sphere indicates that the phase particle is trapped in the potential well, representing that the CBJJ is in the superconducting state, while the red sphere indicates that the particle is escaped from the potential well and thus the CBJJ is in the voltage state. The black-, red-, green-, and blue solid lines represent the potential energy curves for the dc-biased currents being $i_{b}=0, 0.2, 0.5$ and $0.9$, respectively.}
  \label{fig:barrier}
\end{figure}

The dynamics for the macroscopic phase of a CBJJ device can be described by the equation~\cite{instruments5030025}:
$d^2\varphi/d\tau^{2}+\alpha d\varphi/d\tau+\sin\varphi=i_b\,,
\label{eq:current_eq}$
where $\tau=\omega_{J}t$, $i_b=I_b/I_0$, $\alpha=1/(R_{J}C_{J}\omega_{J})$, $\omega_{J}=\sqrt{2eI_0/\hbar C_J}$ is the plasma frequency at zero-bias of the junction. And, $R_J$, $C_J$, and $I_0$ represent the resistance, capacitance, and critical current of the Josephson junction, respectively. As shown in Fig.~\ref{fig:barrier}, the CBJJ describes a ``phase particle'' moving within the potential field~\cite{PhysRevLett.107.217401}
\begin{eqnarray}
U(\varphi)=-E_J[i_{b}\varphi+\cos\varphi]\,,
\label{eq:potential}
\end{eqnarray}
with an ``effective mass'' given by $m^{\ast}=(\hbar/2e)^{2}C_{J}$. Here, $E_J=I_c\Phi_0/2\pi$ and $\Phi_0=h/2e$ represent the Josephson energy and the magnetic flux quantum, respectively. The corresponding potential barrier height is given by~\cite{DElia2023SteppingCT} $\Delta U[i_b(\tau)]=2E_J\left[\sqrt{1-(i_b)^2}-i_b\arccos(i_b)\right]\,\label{eq:barrier}$.
Obviously, if the biased current satisfies that condition $i_b < 1$, then the ``phase particle'' can be trapped in the potential well and the CBJJ should be at a superconducting state with the voltage across the junction being $V = 0$ (marked as the black sphere). With the bias current $i_b$ being increased, the potential for trapping the ``phase particle'', shown in Eq.~\eqref{eq:potential}, becomes skewed, and thus the barrier height decreases. Once the bias current reaches a certain threshold, the ``phase particle'' would escape from the potential well, causing an observable finite voltage state (i.e., $V \neq 0$) across the JJ (denoted by a red sphere). Therefore, microwave photons signal could be regarded effectively as an additionally biased current, which can be detected by probing the change of the threshold described above.   
In a practical CBJJ device, the noise current $i_n(\tau)$ should exist. Also, to implement the rapid response probe, the biased current should be set as time-dependent, i.e., $i_b\rightarrow i_b(\tau)$. Therefore, the dynamics for the phase particle of the practical JJ device biased by a time-dependent current and noise for the detected microwave current could be expressed as 
\begin{eqnarray}
&\frac{\rm d^{2}\varphi}{\rm d\tau^{2}}+\alpha\frac{\rm d\varphi}{\rm d\tau}+\sin\varphi=i_{b}(\tau)+i_{n}(\tau)+i_{s}(\tau).
\label{eq:switch_time}
\end{eqnarray}
where $i_s(\tau) = (I_{MW}/I_0) \sin[(\omega_s/\omega_J) \tau]$ is the microwave signal current expected to be detected. The biased time-dependent current could be simply set as $i_b(\tau) = v\tau$ with $v$ being the dimensionless biased current scan rate and $i_n(\tau)$ refers to the equivalent random noise current satisfying usually the following statistical law~\cite{9919334}:
\begin{eqnarray}
\langle i_{n}(\tau)\rangle=0,\,~
\langle i_{n}(\tau),i_{n}(\tau^{\prime})\rangle=4\alpha D\delta(\tau-\tau^{\prime}).
\end{eqnarray}
Here, $\langle x\rangle$ represents the statistical mean value of the variable $x$, $\delta$ is the Dirac delta function, and $D=2ek_BT/(\hbar I_0)$ is the normalized noise intensity with $T=50$~mK being temperature and $k_B=1.38\times10^{-23}$~J/K the Boltzmann constant. 

\begin{figure}[htbp]
  \centering
  \begin{minipage}{0.25\textwidth}
  \vspace{3pt}
  \centerline{\includegraphics[width=\linewidth]{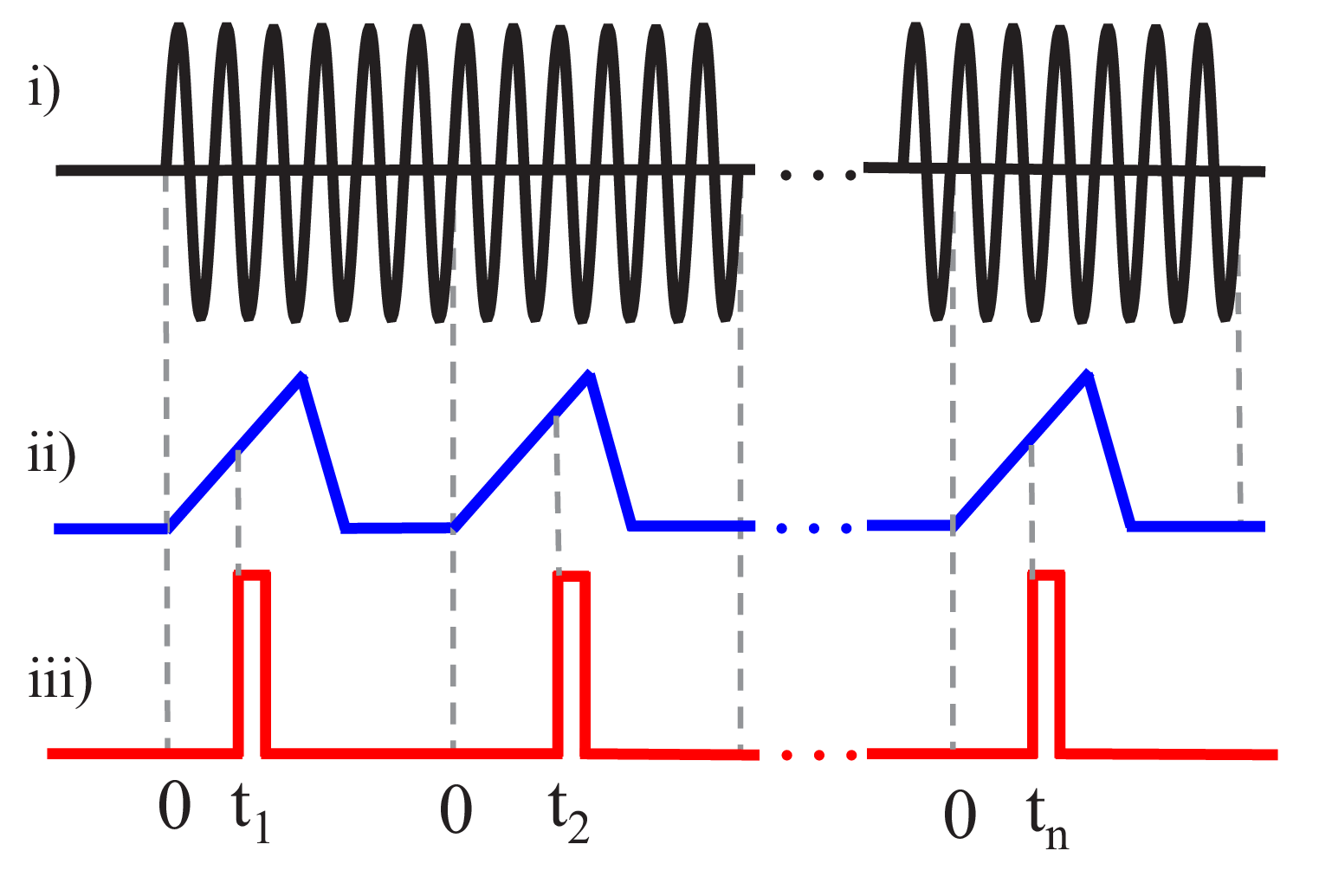}}
  \centerline{(a)}
  \end{minipage}
  \begin{minipage}{0.21\textwidth}
  \centerline{\includegraphics[width=\linewidth]{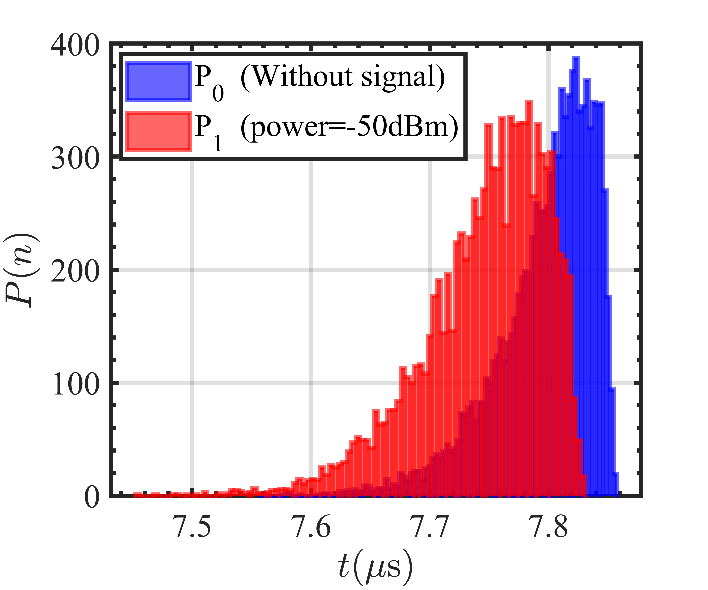}}
  \centerline{(b)}
  \end{minipage}
  \caption{(a) The relevant signals for the detection of the microwave by using the CBJJ device; i) The input microwave signal, ii) the biased triangular wave currents, and iii) the detected voltage switched durations across the JJ at $t_i$, $i=1,\dots,n$. (b) The simulated statistical distributions of the observable $N(= 10^{4})$ times voltage-switched events for microwave detection by using the CBJJ device. Here, $P_0$ and $P_1$ refer to the absence and presence of the input microwave signal (with the power being $-50$~dBm), respectively. Here, the length of the time step is set as $\Delta t=0.02$ and the parameters for the present numerical simulations are set as~\cite{PhysRevE.98.012140}: $\alpha=1.0\times10^{-3}$, $v = 1.0\times10^{-4}$, $D=1.0\times 10^{-3}$, $\omega_s=8$~GHz, $I_0=4.18~\mu A$, $R=100 \Omega$, and $C_J=7.83$~nF, respectively.}
  \label{fig:Impulse&simulation}
\end{figure}
The Josephson threshold detection(JTD) of the applied microwave photon signal $i_s(\tau)$ can be performed as follows.  
First, without the detected microwave signal, we probe the voltage switched event, e.g., the voltage-state switched duration $t$, beginning at the readout pulse be applied to the CBJJ. Here, the readout pulse is just the biased time-dependent current $i_b(\tau)$, shown typically in Fig.~\ref{fig:Impulse&simulation}(a), wherein $t_i, i=1,2,3,...$ is the switched duration of the CBJJ for the $i$th measurement. Due to the existence of the random noise $i_n(\tau)$, the measured switched durations should show a statistical distribution $P_0$ for the repeatedly $N$-times measurements, see e.g., the blue histogram in Fig.~\ref{fig:Impulse&simulation}(b).
Secondly, let us probe the corresponding voltage-switched durations for the successive microwave signal driving $i_s(\tau)$, in the presence of the noise $i_n(\tau)$. After repeating  N times probes, the measured switched durations should show another statistical distribution, e.g.,  the red histogram $P_1$ in Fig.~\ref{fig:Impulse&simulation}(b). 
Thirdly, the detectability of the applied microwave signal can be quantitatively determined by computing the Kumar-Caroll (KC) index~\cite{PhysRevApplied.16.054015}, 
\begin{eqnarray}
d_{\rm KC}=\frac{|\langle t(sw)_1\rangle-\langle \textit{t}(sw)_0\rangle|}
{\sqrt{\frac{1}{2}(\sigma_1^2+\sigma_0^2)}}\,,\label{eq:d_KC}
\end{eqnarray}
to implement the discrimination between the two measured switched duration distributions. Above, $\langle t(sw)_j\rangle=(1/N)\sum_{i=1}^{N}t_{i}$, $j = 0,1$, represents the statistical average of the measured switched durations with the variance 
$\sigma_j^2=[1/N(N-1)]\sum_{i=1}^N(t_i-\langle t(sw)_j\rangle)^2\,(j=0,1)$ for the $P_j$-distribution. 
Obviously, $d_{\rm KC}\geq 1$ refers that two statistical distributions are realizable discrimination~\cite{9555447}, and the larger $d_{\rm KC}$-value corresponds to the better differentiation. Physically, the minimal value, i.e., min[$d_{\rm KC}$], corresponds to the noise equivalent power (NEP) of the device~\cite{PhysRevB.111.024501}. 

Given the Josephson threshold detector had been utilized to demonstrate the sensitive detection of infrared photons~\cite{ali_josephson_2024}, below we show how such a detector can also be used to implement the detection of the microwave photons, treated here as a coherent microwave current $i_b(\tau)$ applied to the CBJJ device, alternatively. 
The JJ device used here is an Al/AlOx/Al-structure, which is fabricated by using the oblique-angle electron beam evaporation and laser direct-write lithography, see Appendix A for details. Fig.~\ref{fig:JJ_observe}(a) and Fig.~\ref{fig:JJ_observe}(b)-(c) show the configuration of the designed device and the microphotograph of the fabricated device, respectively. The I-V curve and the switched current distribution of the fabricated JJ device were measured at 50~mK cryogenic environment provided by a Leiden Cryogen-free dilution refrigerator, see Appendix B for the details of the measurement circuit. The relevant results are shown in Fig.~\ref{fig:IV_fit}, and fitted by the relevant numerical simulations, based on the usual RCSJ model and the switched current distribution formula~\cite{2008_Cui,PhysRevB.28.1268,PhysRevLett.46.211}$; P(I)=-dN/dI=\Gamma(I)(dI/dt)^{-1}\times e^{-(dI/dt)^{-1}\int^{I}_{0}\Gamma(I')dI'}$. As a consequence, the physical parameters of the measured JJ can be extracted as~\cite{han_experimentally_2021,PhysRevResearch.6.013236}; $C_J\sim 1$~pF (the junction capacitance), $R_J\sim 480\Omega$, and $I_0\sim 1.4\times 10^{-7}$~A (its critical current). Consequently, the plasma frequency and damping coefficient of the JJ can be numerically estimated as $\omega_J/2\pi\sim 3.3$ GHz and $\alpha\sim 0.101$, respectively.   
\begin{figure}[htbp]
  \centering
  \includegraphics[width=0.9\linewidth]{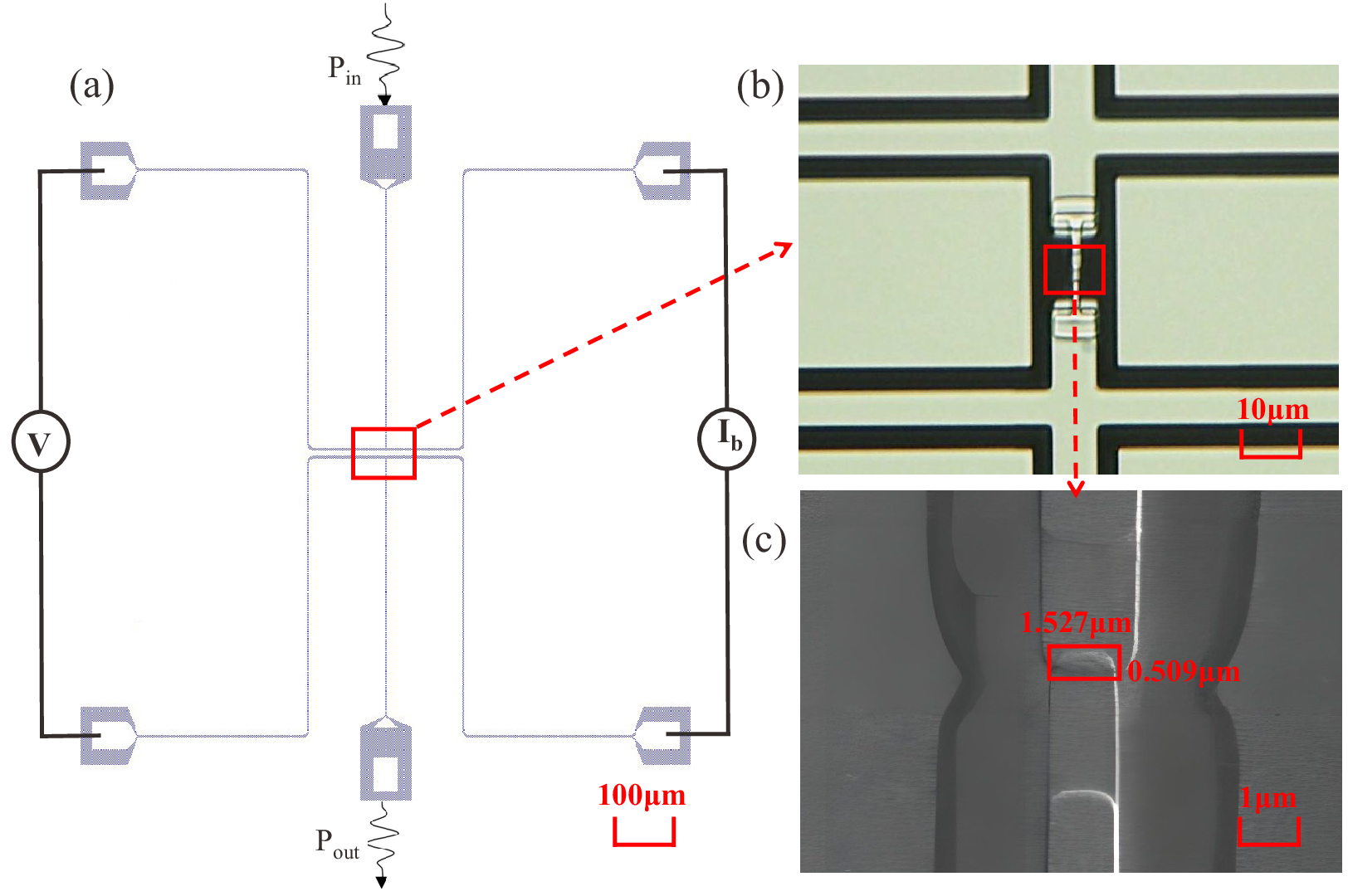}
  \caption{(a) Schematic diagram of the CBJJ detecting a microwave signal. The corner ports are used to apply a bias current $I_b$ to the junction and to detect the voltage $V$ across it, while the top and bottom ports are used for input and output of microwave signals. (b) Image of the junction region captured with a high-magnification microscope. (c) Image of the junction region obtained using a scanning electron microscope.}
  \label{fig:JJ_observe}
\end{figure}
\begin{figure}[htbp]
  \centering
  \begin{minipage}{0.23\textwidth}
  \vspace{3pt}
  \centerline{\includegraphics[width=\linewidth]{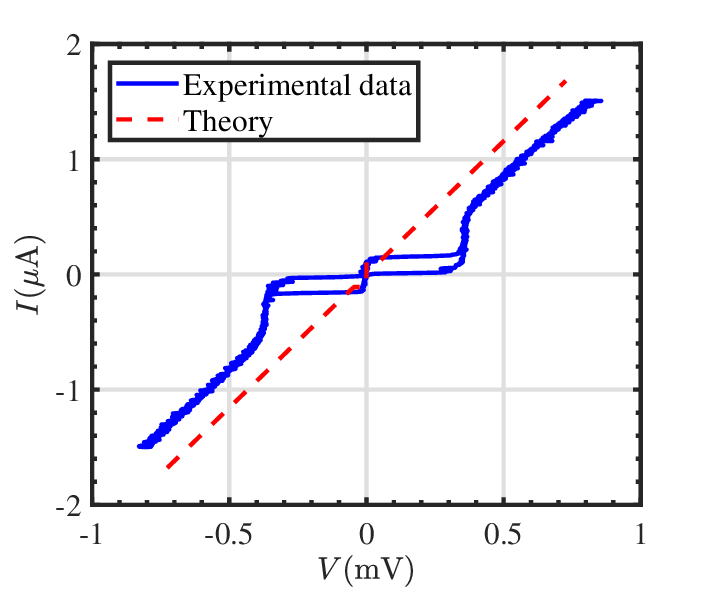}}
  \centerline{(a)}
  \end{minipage}
  \begin{minipage}{0.23\textwidth}
  \centerline{\includegraphics[width=\linewidth]{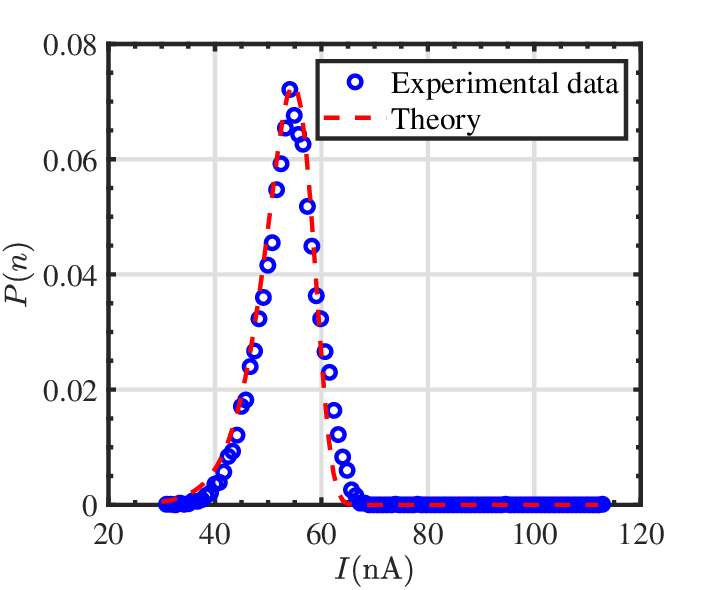}}
  \centerline{(b)}
  \end{minipage}
  \caption{(a) Josephson junction IV curve: the blue solid line represents the experimental data, while the red dashed line represents the theoretical simulation. (b) Experimental and theoretical fitting results of the Josephson junction switched current. Histograms showing the distribution of the switched current obtained from 10,000 junction switched events driven by a triangular wave. The blue circles represent experimental data, and the red dashed line represents the theoretically fitted distribution.}
  \label{fig:IV_fit}
\end{figure}
For the implementation of the sensitive microwave signal detection by using the threshold response of the CBJJ device, we first measure the microwave signal transporting attenuation along the measurement circuit. It is seen clearly from Appendix B that, the microwave transporting circuit of the JJ device, without any current bias, exists about $-2$~dB attenuation, if the microwave frequency is set around $\omega_s/2\pi\sim 5$~GHz. The experimental results for the measured $S_{21}$-parameters shown in Fig.~\ref{fig:JJe_life}(a) indicates that the relatively weak and stable attenuation revealed for the microwave signals being set in the 4.5-5.5 GHz frequency band. In contrast, microwave signals in other frequency bands are strongly and unstable attenuated, which is probably due to either the strong absorption by the JJ itself (or its impurities) or the strong attenuation of the measurement circuit. Therefore, to minimize the potential physical perturbations, we perfectly measure the microwave signals' response at $\sim 5$~GHz first, and let those for the microwave signals in the other frequency bands be investigated later.

Following the method described above in the fourth paragraph, the microwave signal could be detected by applying it to the JJ and then probing its switched duration of voltage pulse across the JJ. The measured voltage switched duration distributions are shown in Fig.~\ref{fig:JJe_life}(b), for the typical microwave signal powers. Here, for each of the microwave power signals, the detection is repeated by $10^4$ times.
\begin{figure}[htbp]
  \centering
  \begin{minipage}{0.23\textwidth}
  \vspace{3pt}
  \centerline{\includegraphics[width=\linewidth]{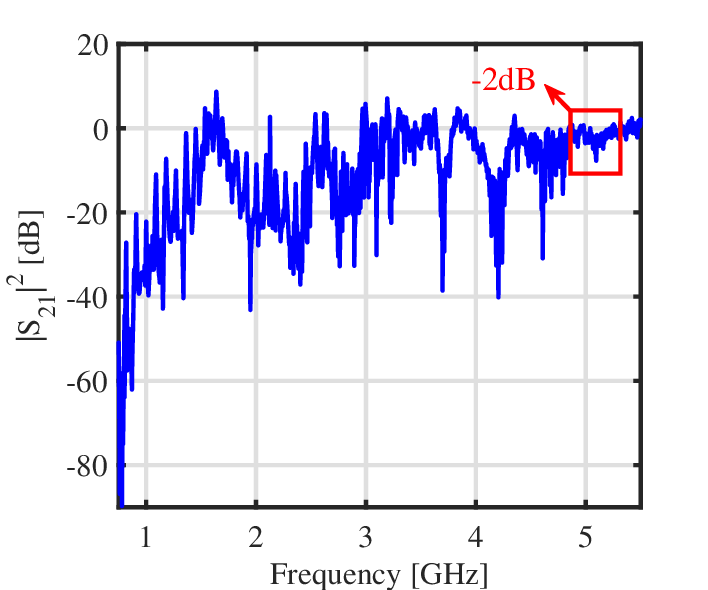}}
  \centerline{(a)}
  \end{minipage}
  \begin{minipage}{0.23\textwidth}
  \centerline{\includegraphics[width=\linewidth]{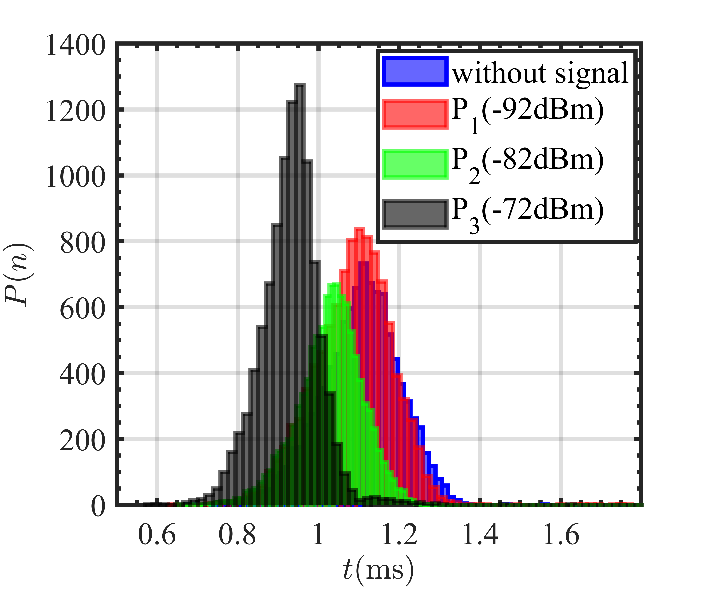}}
  \centerline{(b)}
  \end{minipage}
  \caption{(a) $S_{21}$ data for the Josephson junction at zero bias. The red box highlights the 4.5~GHz to 5.5~GHz band, which exhibits an averaged attenuation of -2~dBm. (b) Statistics of the switched duration of the Josephson junction driven by microwave signals with varying powers at a frequency of $\sim 5$~GHz. The blue distribution represents the switched duration distribution in the noise background without signal input, while the black, red, and green histograms correspond to the switched duration distributions at powers of -72~dBm, -82~dBm, and -92~dBm, respectively.}
  \label{fig:JJe_life}
\end{figure}
To differentiate them from the distribution without any microwave signal, we calculate the relevant $d_{\rm KC}$ indexes and list them in Table.~\ref{table:KC} for comparison. 
\begin{table}[htbp]
\caption{\label{table:KC} The calculated $d_{\rm KC}$ indexes between the measured statistical distributions of voltage switched durations of the JJ driven by the input microwaves with different powers and that for the JJ without any microwave driving.}
\begin{ruledtabular}
\begin{tabular}{cccc}
& $P_1$ & $P_2$ & $P_3$\\
\hline
power & $\rm -92~dBm$ & $\rm -82~dBm$ & $\rm -72~dBm$\\
\hline
$d_{\rm KC}$ & 0.1813 & 0.8498 & 2.0003\\
\end{tabular}
\end{ruledtabular}
\end{table}
Obviously, with the increase of the power of the incident microwave signal, the voltage switched duration distribution is the stronger distinguishability from that switched by the pure noise drivings. Following the Ref.~\onlinecite{PhysRevB.111.024501}, the minimum $d_{\rm KC}$ index of the present Josephson device is quantitatively estimated as $\min[d_{\rm KC}]=0.11$, corresponding to the noise-equivalent-power (NEP), i.e., the minimum detectable signal power of the present detector. It is seen from Table.~\ref{table:KC} that, the signal around $5$~GHz with the power $-92$~dBm (i.e., the microwave signal includes about $3\times10^{7}$~photons/ms) is still far from its minimum detectable signal power, as the corresponding $d_{\rm KC}$ is still manifestly larger than its minimal value. This implies that the achievable detection sensitivity for the $\sim 5$GHz microwave photons had not yet arrived. We argue that there is still about two orders of magnitude of room for improvement. Physically, this result is reasonable, as the presently detected microwave signal frequency is $\sim 5$~GHz, which is obviously deviated from the plasma frequency ($\sim 3.3$~GHz) of the device and thus possesses the weakest attenuation. The larger attenuation corresponds to the stronger absorption.  
This is consistent with the $S_{21}$-parameter measurement data shown in Fig.~\ref{fig:JJe_life}(a), i.e., the transport attenuation is almost the weakest for the frequency of the transporting being set as $\sim 5$~GHz, as the weaker attenuation implies the weaker photon-detector coupling, yielding the lower detection sensitivity.

\begin{figure}[htbp]
  \centering
  \begin{minipage}{0.23\textwidth}
  \vspace{3pt}
  \centerline{\includegraphics[width=\linewidth]{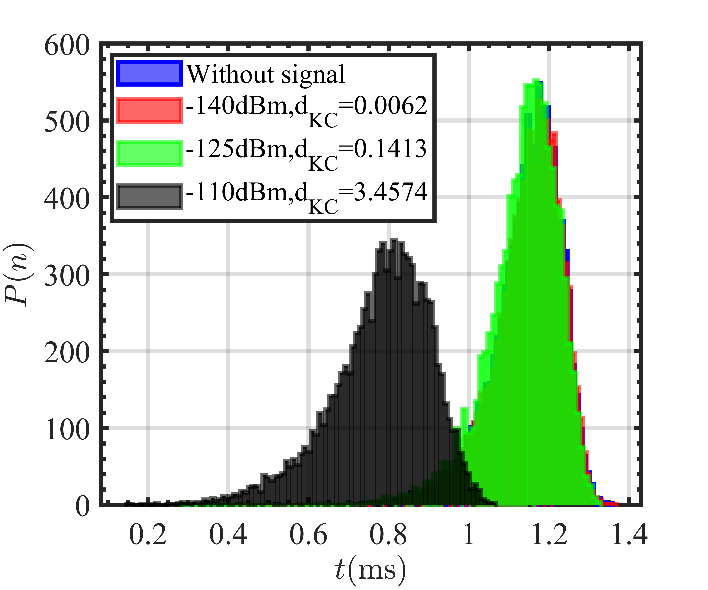}}
  \centerline{(a)}
  \end{minipage}
  \begin{minipage}{0.23\textwidth}
  \centerline{\includegraphics[width=\linewidth]{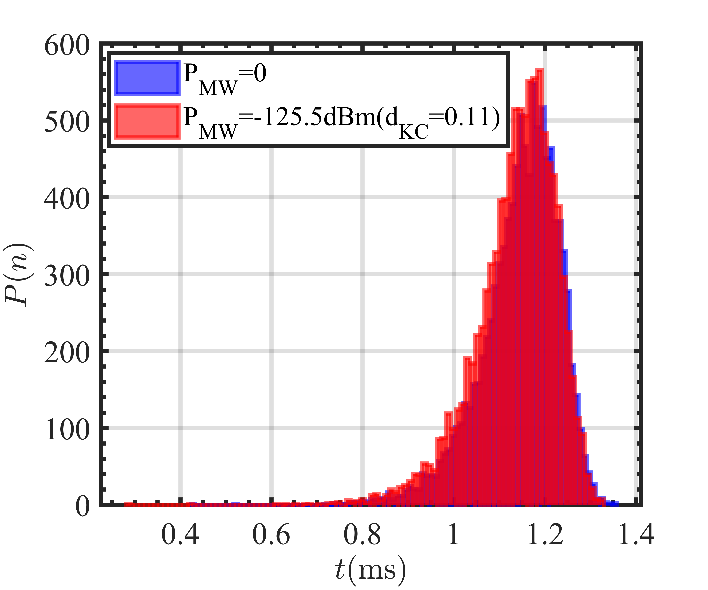}}
  \centerline{(b)}
  \end{minipage}
  \caption{Simulation of the voltage switched duration $t$ (ms) for the CBJJ driven by the microwave photons whose frequency is set at the plasma frequency~$\omega_J/2\pi\approx 3.3$GHz of the JJ. The distribution is obtained by solving Eq.~\eqref{eq:switch_time} for $10^4$ times. (a) The different KC indexes for the corresponding microwave signal power, and (b) the signal power approaches the NEP of the device and its $d_{KC}$ index. Here, the relevant experimental parameters are set as; $I_0=1.4\times10^{-7}$~A, $R=480~\Omega$, $C=1$~pF, for the JJ and $v=56.5\times10^{-6}$~(A/s) for the readout current, at $T=50$~mK.}
  \label{fig:simulation_wJ}
\end{figure}

Physically, high detection sensitivity implies a strong interaction between the microwave photons and the JTD, which thus yields the strong attenuation of the traveling microwave photons. This implies that the achievable detection sensitivity of the present JTD could be significantly improved if it is applied to implement the detection of microwave photons with the frequency around the plasma frequency of the fabricated JJ. This is because, the microwave photons near this frequency can resonantly excite the Josephson junction and thus facilitate the sensitive voltage switching, provided that complex noise factors are effectively minimized. To numerically verify such a speculation, we solve Eq.~\eqref{eq:switch_time} with $\omega_s/(2\pi)=\omega_J/(2\pi)=3.3$~GHz for the present JTD. In the absence/presence of the microwave driving, the simulative switched duration distributions, induced by the applied microwave current (with the frequency of $3.3$~GHz), are shown in Fig.~\ref{fig:JJe_life}(a) for the voltage switched duration of the JJ. The simulated results are shown specifically in Fig.~\ref{fig:simulation_wJ} for different powers of the incident microwave signals. Therefore, it is argued that, with the present JTD device (with the min$[d_{\rm KC}]=0.11$), the detection of the microwave photons of $\sim 3.3$GHz with the power of $\sim -125.5$~dBm is possible. This argument will be checked in our future experiments.        

In summary, we demonstrated the feasibility of using the fabricated JTD to implement the detection of weak microwave signals. The present JTD device consists of an Al/AlOx/A JJ and the readout driving generated by its biased triangular wave currents. By using 10,000 repeated measurements of the voltage switched duration of the JTD, with and without the detected microwave signal inputs, and then calculating the $d_{\rm KC}$ index of the measured distributions, we showed that the JTD demonstrated here can be applied to detect the microwave signal (whose frequency is $\sim 5$~GHz) with the sensitivity of being $\sim -92$~dBm. It is argued, by the further numerical simulations,  that the present JTD, generated by a JJ with the critical current of $I_0 \sim 0.14~\mu$A, the junction capacitance of $C_J \sim 1$ pF, and the junction resistance of $R_J \sim 480~\Omega$, could be applied to implement the highly sensitive detection of the microwave signals, whose frequency is near the plasma frequency (i.e., $\sim 3.3$~GHz)  of the JJ, and the achievable detection sensitivity can be enhanced as $-125.5$~dBm. This possibility will be tested by future experiments.

Although the sensitivity demonstrated here for the detection of microwave photons (at $\sim 5$~GHz) is manifestly lower than the predicted NEP of the device and those speculated in Refs.~\onlinecite{REVIN2020960,PANKRATOV2022582,pankratov_towards_2022}, we argued that the achievable detection sensitivity of the generated JTD could be significantly improved; i) for the detection of microwave photons with the frequency being around the plasma frequency of the device, ii) by further optimizing the physical parameters of the device (e.g., minimize the number of impurities in the junction) and the measurement circuit of the microwave transports~\cite{PhysRevB.111.024501,DElia2023SteppingCT}. Anyway, we believe that the experimental JTD possesses the potential to realize the desirably sensitive microwave single-photon detection.  


This work was partially supported by the National Key Research and Development Program of China (NKRDC) under Grant No. 2021YFA0718803 and the National Natural Science Foundation of China (NSFC) under Grant No. 11974290.

\section*{AUTHOR DECLARATIONS}

\subsection*{Conflict of Interest}

The authors have no conflicts to disclose.

\section*{DATA AVAILABILITY}

The data that support the findings of this study are available from the corresponding authors upon reasonable request.

\nocite{*}
\bibliographystyle{aipnum4-1}
\bibliography{aipsamp}

\begin{thebibliography}{39}%
\makeatletter
\providecommand \@ifxundefined [1]{%
 \@ifx{#1\undefined}
}%
\providecommand \@ifnum [1]{%
 \ifnum #1\expandafter \@firstoftwo
 \else \expandafter \@secondoftwo
 \fi
}%
\providecommand \@ifx [1]{%
 \ifx #1\expandafter \@firstoftwo
 \else \expandafter \@secondoftwo
 \fi
}%
\providecommand \natexlab [1]{#1}%
\providecommand \enquote  [1]{``#1''}%
\providecommand \bibnamefont  [1]{#1}%
\providecommand \bibfnamefont [1]{#1}%
\providecommand \citenamefont [1]{#1}%
\providecommand \href@noop [0]{\@secondoftwo}%
\providecommand \href [0]{\begingroup \@sanitize@url \@href}%
\providecommand \@href[1]{\@@startlink{#1}\@@href}%
\providecommand \@@href[1]{\endgroup#1\@@endlink}%
\providecommand \@sanitize@url [0]{\catcode `\\12\catcode `\$12\catcode
  `\&12\catcode `\#12\catcode `\^12\catcode `\_12\catcode `\%12\relax}%
\providecommand \@@startlink[1]{}%
\providecommand \@@endlink[0]{}%
\providecommand \url  [0]{\begingroup\@sanitize@url \@url }%
\providecommand \@url [1]{\endgroup\@href {#1}{\urlprefix }}%
\providecommand \urlprefix  [0]{URL }%
\providecommand \Eprint [0]{\href }%
\providecommand \doibase [0]{http://dx.doi.org/}%
\providecommand \selectlanguage [0]{\@gobble}%
\providecommand \bibinfo  [0]{\@secondoftwo}%
\providecommand \bibfield  [0]{\@secondoftwo}%
\providecommand \translation [1]{[#1]}%
\providecommand \BibitemOpen [0]{}%
\providecommand \bibitemStop [0]{}%
\providecommand \bibitemNoStop [0]{.\EOS\space}%
\providecommand \EOS [0]{\spacefactor3000\relax}%
\providecommand \BibitemShut  [1]{\csname bibitem#1\endcsname}%
\let\auto@bib@innerbib\@empty
\bibitem [{\citenamefont {Kok}\ \emph {et~al.}(2007)\citenamefont {Kok},
  \citenamefont {Munro}, \citenamefont {Nemoto}, \citenamefont {Ralph},
  \citenamefont {Dowling},\ and\ \citenamefont {Milburn}}]{RevModPhys.79.135}%
  \BibitemOpen
  \bibfield  {author} {\bibinfo {author} {\bibfnamefont {P.}~\bibnamefont
  {Kok}}, \bibinfo {author} {\bibfnamefont {W.~J.}\ \bibnamefont {Munro}},
  \bibinfo {author} {\bibfnamefont {K.}~\bibnamefont {Nemoto}}, \bibinfo
  {author} {\bibfnamefont {T.~C.}\ \bibnamefont {Ralph}}, \bibinfo {author}
  {\bibfnamefont {J.~P.}\ \bibnamefont {Dowling}}, \ and\ \bibinfo {author}
  {\bibfnamefont {G.~J.}\ \bibnamefont {Milburn}},\ }\href {\doibase
  10.1103/RevModPhys.79.135} {\bibfield  {journal} {\bibinfo  {journal} {Rev.
  Mod. Phys.}\ }\textbf {\bibinfo {volume} {79}},\ \bibinfo {pages} {135}
  (\bibinfo {year} {2007})}\BibitemShut {NoStop}%
\bibitem [{\citenamefont {Hadfield}(2009)}]{hadfield_single-photon_2009}%
  \BibitemOpen
  \bibfield  {author} {\bibinfo {author} {\bibfnamefont {R.~H.}\ \bibnamefont
  {Hadfield}},\ }\href {\doibase 10.1038/nphoton.2009.230} {\bibfield
  {journal} {\bibinfo  {journal} {Nat. Photonics}\ }\textbf {\bibinfo {volume}
  {3}},\ \bibinfo {pages} {696} (\bibinfo {year} {2009})}\BibitemShut {NoStop}%
\bibitem [{\citenamefont {Guo}\ \emph {et~al.}(2022)\citenamefont {Guo},
  \citenamefont {Yu}, \citenamefont {Wei}, \citenamefont {Jin}, \citenamefont
  {Chen}, \citenamefont {Li}, \citenamefont {Zhang},\ and\ \citenamefont
  {Zhou}}]{GUO20222291}%
  \BibitemOpen
  \bibfield  {author} {\bibinfo {author} {\bibfnamefont {X.}~\bibnamefont
  {Guo}}, \bibinfo {author} {\bibfnamefont {Z.}~\bibnamefont {Yu}}, \bibinfo
  {author} {\bibfnamefont {F.}~\bibnamefont {Wei}}, \bibinfo {author}
  {\bibfnamefont {S.}~\bibnamefont {Jin}}, \bibinfo {author} {\bibfnamefont
  {X.}~\bibnamefont {Chen}}, \bibinfo {author} {\bibfnamefont {X.}~\bibnamefont
  {Li}}, \bibinfo {author} {\bibfnamefont {X.}~\bibnamefont {Zhang}}, \ and\
  \bibinfo {author} {\bibfnamefont {X.}~\bibnamefont {Zhou}},\ }\href {\doibase
  https://doi.org/10.1016/j.scib.2022.10.027} {\bibfield  {journal} {\bibinfo
  {journal} {Sci. Bull.}\ }\textbf {\bibinfo {volume} {67}},\ \bibinfo {pages}
  {2291} (\bibinfo {year} {2022})}\BibitemShut {NoStop}%
\bibitem [{\citenamefont {Feng}, \citenamefont {Zhang},\ and\ \citenamefont
  {Wei}(2024)}]{PhysRevLett.132.220801}%
  \BibitemOpen
  \bibfield  {author} {\bibinfo {author} {\bibfnamefont {X.~N.}\ \bibnamefont
  {Feng}}, \bibinfo {author} {\bibfnamefont {M.}~\bibnamefont {Zhang}}, \ and\
  \bibinfo {author} {\bibfnamefont {L.~F.}\ \bibnamefont {Wei}},\ }\href
  {\doibase 10.1103/PhysRevLett.132.220801} {\bibfield  {journal} {\bibinfo
  {journal} {Phys. Rev. Lett.}\ }\textbf {\bibinfo {volume} {132}},\ \bibinfo
  {pages} {220801} (\bibinfo {year} {2024})}\BibitemShut {NoStop}%
\bibitem [{\citenamefont {Luong}, \citenamefont {Balaji},\ and\ \citenamefont
  {Rajan}(2023)}]{10209149}%
  \BibitemOpen
  \bibfield  {author} {\bibinfo {author} {\bibfnamefont {D.}~\bibnamefont
  {Luong}}, \bibinfo {author} {\bibfnamefont {B.}~\bibnamefont {Balaji}}, \
  and\ \bibinfo {author} {\bibfnamefont {S.}~\bibnamefont {Rajan}},\ }\href
  {\doibase 10.1109/MMM.2023.3284765} {\bibfield  {journal} {\bibinfo
  {journal} {IEEE Microw. Mag.}\ }\textbf {\bibinfo {volume} {24}},\ \bibinfo
  {pages} {61} (\bibinfo {year} {2023})}\BibitemShut {NoStop}%
\bibitem [{\citenamefont {Zhang}\ \emph {et~al.}(2021)\citenamefont {Zhang},
  \citenamefont {Chen}, \citenamefont {Guan}, \citenamefont {Jin},
  \citenamefont {Wang}, \citenamefont {Zhang}, \citenamefont {Tu},
  \citenamefont {Zhao}, \citenamefont {Jia}, \citenamefont {Kang},
  \citenamefont {Chen},\ and\ \citenamefont {Wu}}]{zhang_research_2021}%
  \BibitemOpen
  \bibfield  {author} {\bibinfo {author} {\bibfnamefont {B.}~\bibnamefont
  {Zhang}}, \bibinfo {author} {\bibfnamefont {Q.}~\bibnamefont {Chen}},
  \bibinfo {author} {\bibfnamefont {Y.-Q.}\ \bibnamefont {Guan}}, \bibinfo
  {author} {\bibfnamefont {F.-F.}\ \bibnamefont {Jin}}, \bibinfo {author}
  {\bibfnamefont {H.}~\bibnamefont {Wang}}, \bibinfo {author} {\bibfnamefont
  {L.-B.}\ \bibnamefont {Zhang}}, \bibinfo {author} {\bibfnamefont {X.-C.}\
  \bibnamefont {Tu}}, \bibinfo {author} {\bibfnamefont {Q.-Y.}\ \bibnamefont
  {Zhao}}, \bibinfo {author} {\bibfnamefont {X.-Q.}\ \bibnamefont {Jia}},
  \bibinfo {author} {\bibfnamefont {L.}~\bibnamefont {Kang}}, \bibinfo {author}
  {\bibfnamefont {J.}~\bibnamefont {Chen}}, \ and\ \bibinfo {author}
  {\bibfnamefont {P.-H.}\ \bibnamefont {Wu}},\ }\href {\doibase
  10.7498/aps.70.20210652} {\bibfield  {journal} {\bibinfo  {journal} {Acta
  Phys. Sin.}\ }\textbf {\bibinfo {volume} {70}},\ \bibinfo {pages} {198501}
  (\bibinfo {year} {2021})}\BibitemShut {NoStop}%
\bibitem [{\citenamefont {Friedrich}\ \emph {et~al.}(2022)\citenamefont
  {Friedrich}, \citenamefont {Marino}, \citenamefont {Ponce}, \citenamefont
  {Carpenter}, \citenamefont {Kim}, \citenamefont {Drury}, \citenamefont
  {Drake}, \citenamefont {Bray}, \citenamefont {Fretwell}, \citenamefont
  {Leach}, \citenamefont {Harris}, \citenamefont {Warburton}, \citenamefont
  {Hall},\ and\ \citenamefont {Cantor}}]{friedrich_characterization_2022}%
  \BibitemOpen
  \bibfield  {author} {\bibinfo {author} {\bibfnamefont {S.}~\bibnamefont
  {Friedrich}}, \bibinfo {author} {\bibfnamefont {A.}~\bibnamefont {Marino}},
  \bibinfo {author} {\bibfnamefont {F.}~\bibnamefont {Ponce}}, \bibinfo
  {author} {\bibfnamefont {M.~F.}\ \bibnamefont {Carpenter}}, \bibinfo {author}
  {\bibfnamefont {G.-B.}\ \bibnamefont {Kim}}, \bibinfo {author} {\bibfnamefont
  {O.~B.}\ \bibnamefont {Drury}}, \bibinfo {author} {\bibfnamefont
  {J.}~\bibnamefont {Drake}}, \bibinfo {author} {\bibfnamefont {C.~R.}\
  \bibnamefont {Bray}}, \bibinfo {author} {\bibfnamefont {S.}~\bibnamefont
  {Fretwell}}, \bibinfo {author} {\bibfnamefont {K.~G.}\ \bibnamefont {Leach}},
  \bibinfo {author} {\bibfnamefont {J.}~\bibnamefont {Harris}}, \bibinfo
  {author} {\bibfnamefont {W.~K.}\ \bibnamefont {Warburton}}, \bibinfo {author}
  {\bibfnamefont {J.~A.}\ \bibnamefont {Hall}}, \ and\ \bibinfo {author}
  {\bibfnamefont {R.}~\bibnamefont {Cantor}},\ }\href {\doibase
  10.1007/s10909-022-02825-6} {\bibfield  {journal} {\bibinfo  {journal} {J.
  Low Temp. Phys.}\ }\textbf {\bibinfo {volume} {209}},\ \bibinfo {pages}
  {1063} (\bibinfo {year} {2022})}\BibitemShut {NoStop}%
\bibitem [{\citenamefont {Dai}\ \emph {et~al.}(2025)\citenamefont {Dai},
  \citenamefont {Wang}, \citenamefont {Wang}, \citenamefont {Mai},
  \citenamefont {Shi}, \citenamefont {Wang}, \citenamefont {Jia}, \citenamefont
  {Liu}, \citenamefont {He}, \citenamefont {Dai}, \citenamefont {Ouyang},
  \citenamefont {Chai}, \citenamefont {Wei}, \citenamefont {Zhang},
  \citenamefont {Zhong}, \citenamefont {Guo}, \citenamefont {Liu},\ and\
  \citenamefont {Yu}}]{10.1063/5.0234649}%
  \BibitemOpen
  \bibfield  {author} {\bibinfo {author} {\bibfnamefont {X.}~\bibnamefont
  {Dai}}, \bibinfo {author} {\bibfnamefont {H.}~\bibnamefont {Wang}}, \bibinfo
  {author} {\bibfnamefont {Y.}~\bibnamefont {Wang}}, \bibinfo {author}
  {\bibfnamefont {Z.}~\bibnamefont {Mai}}, \bibinfo {author} {\bibfnamefont
  {Z.}~\bibnamefont {Shi}}, \bibinfo {author} {\bibfnamefont {Y.-F.}\
  \bibnamefont {Wang}}, \bibinfo {author} {\bibfnamefont {H.}~\bibnamefont
  {Jia}}, \bibinfo {author} {\bibfnamefont {J.}~\bibnamefont {Liu}}, \bibinfo
  {author} {\bibfnamefont {Q.}~\bibnamefont {He}}, \bibinfo {author}
  {\bibfnamefont {M.}~\bibnamefont {Dai}}, \bibinfo {author} {\bibfnamefont
  {P.}~\bibnamefont {Ouyang}}, \bibinfo {author} {\bibfnamefont
  {Y.}~\bibnamefont {Chai}}, \bibinfo {author} {\bibfnamefont {L.-F.}\
  \bibnamefont {Wei}}, \bibinfo {author} {\bibfnamefont {L.}~\bibnamefont
  {Zhang}}, \bibinfo {author} {\bibfnamefont {Y.}~\bibnamefont {Zhong}},
  \bibinfo {author} {\bibfnamefont {W.}~\bibnamefont {Guo}}, \bibinfo {author}
  {\bibfnamefont {S.}~\bibnamefont {Liu}}, \ and\ \bibinfo {author}
  {\bibfnamefont {D.}~\bibnamefont {Yu}},\ }\href {\doibase 10.1063/5.0234649}
  {\bibfield  {journal} {\bibinfo  {journal} {Appl. Phys. Lett.}\ }\textbf
  {\bibinfo {volume} {126}},\ \bibinfo {pages} {012602} (\bibinfo {year}
  {2025})}\BibitemShut {NoStop}%
\bibitem [{\citenamefont {FUKUDA}(2019)}]{Daiji_FUKUDA20192018SDI0001}%
  \BibitemOpen
  \bibfield  {author} {\bibinfo {author} {\bibfnamefont {D.}~\bibnamefont
  {FUKUDA}},\ }\href {\doibase 10.1587/transele.2018SDI0001} {\bibfield
  {journal} {\bibinfo  {journal} {IEICE Transactions on Electronics}\ }\textbf
  {\bibinfo {volume} {E102.C}},\ \bibinfo {pages} {230} (\bibinfo {year}
  {2019})}\BibitemShut {NoStop}%
\bibitem [{\citenamefont {Pankratov}\ \emph
  {et~al.}(2022{\natexlab{a}})\citenamefont {Pankratov}, \citenamefont
  {Gordeeva}, \citenamefont {Revin}, \citenamefont {Ladeynov}, \citenamefont
  {Yablokov},\ and\ \citenamefont {Kuzmin}}]{PANKRATOV2022582}%
  \BibitemOpen
  \bibfield  {author} {\bibinfo {author} {\bibfnamefont {A.~L.}\ \bibnamefont
  {Pankratov}}, \bibinfo {author} {\bibfnamefont {A.~V.}\ \bibnamefont
  {Gordeeva}}, \bibinfo {author} {\bibfnamefont {L.~S.}\ \bibnamefont {Revin}},
  \bibinfo {author} {\bibfnamefont {D.~A.}\ \bibnamefont {Ladeynov}}, \bibinfo
  {author} {\bibfnamefont {A.~A.}\ \bibnamefont {Yablokov}}, \ and\ \bibinfo
  {author} {\bibfnamefont {L.~S.}\ \bibnamefont {Kuzmin}},\ }\href {\doibase
  https://doi.org/10.3762/bjnano.13.50} {\bibfield  {journal} {\bibinfo
  {journal} {Beilstein J. Nanotechnol.}\ }\textbf {\bibinfo {volume} {13}},\
  \bibinfo {pages} {582} (\bibinfo {year} {2022}{\natexlab{a}})}\BibitemShut
  {NoStop}%
\bibitem [{\citenamefont {Beck}(2013)}]{PhysRevLett.111.231801}%
  \BibitemOpen
  \bibfield  {author} {\bibinfo {author} {\bibfnamefont {C.}~\bibnamefont
  {Beck}},\ }\href {\doibase 10.1103/PhysRevLett.111.231801} {\bibfield
  {journal} {\bibinfo  {journal} {Phys. Rev. Lett.}\ }\textbf {\bibinfo
  {volume} {111}},\ \bibinfo {pages} {231801} (\bibinfo {year}
  {2013})}\BibitemShut {NoStop}%
\bibitem [{\citenamefont {Co}\ \emph {et~al.}(2019)\citenamefont {Co},
  \citenamefont {Pierce}, \citenamefont {Zhang},\ and\ \citenamefont
  {Zhao}}]{PhysRevD.99.075002}%
  \BibitemOpen
  \bibfield  {author} {\bibinfo {author} {\bibfnamefont {R.~T.}\ \bibnamefont
  {Co}}, \bibinfo {author} {\bibfnamefont {A.}~\bibnamefont {Pierce}}, \bibinfo
  {author} {\bibfnamefont {Z.}~\bibnamefont {Zhang}}, \ and\ \bibinfo {author}
  {\bibfnamefont {Y.}~\bibnamefont {Zhao}},\ }\href {\doibase
  10.1103/PhysRevD.99.075002} {\bibfield  {journal} {\bibinfo  {journal} {Phys.
  Rev. D}\ }\textbf {\bibinfo {volume} {99}},\ \bibinfo {pages} {075002}
  (\bibinfo {year} {2019})}\BibitemShut {NoStop}%
\bibitem [{\citenamefont {Zheng}\ \emph {et~al.}(2018)\citenamefont {Zheng},
  \citenamefont {Wei}, \citenamefont {Wen},\ and\ \citenamefont
  {Li}}]{PhysRevD.98.064028}%
  \BibitemOpen
  \bibfield  {author} {\bibinfo {author} {\bibfnamefont {H.}~\bibnamefont
  {Zheng}}, \bibinfo {author} {\bibfnamefont {L.~F.}\ \bibnamefont {Wei}},
  \bibinfo {author} {\bibfnamefont {H.}~\bibnamefont {Wen}}, \ and\ \bibinfo
  {author} {\bibfnamefont {F.~Y.}\ \bibnamefont {Li}},\ }\href {\doibase
  10.1103/PhysRevD.98.064028} {\bibfield  {journal} {\bibinfo  {journal} {Phys.
  Rev. D}\ }\textbf {\bibinfo {volume} {98}},\ \bibinfo {pages} {064028}
  (\bibinfo {year} {2018})}\BibitemShut {NoStop}%
\bibitem [{\citenamefont {Zheng}\ and\ \citenamefont
  {Wei}(2022)}]{PhysRevD.106.104003}%
  \BibitemOpen
  \bibfield  {author} {\bibinfo {author} {\bibfnamefont {H.}~\bibnamefont
  {Zheng}}\ and\ \bibinfo {author} {\bibfnamefont {L.~F.}\ \bibnamefont
  {Wei}},\ }\href {\doibase 10.1103/PhysRevD.106.104003} {\bibfield  {journal}
  {\bibinfo  {journal} {Phys. Rev. D}\ }\textbf {\bibinfo {volume} {106}},\
  \bibinfo {pages} {104003} (\bibinfo {year} {2022})}\BibitemShut {NoStop}%
\bibitem [{\citenamefont {Layne}(2014)}]{layne2014receiver}%
  \BibitemOpen
  \bibfield  {author} {\bibinfo {author} {\bibfnamefont {D.}~\bibnamefont
  {Layne}},\ }\href
  {https://www.highfrequencyelectronics.com/index.php?option=com_content&view=article&id=553:receiver-sensitivity-and-equivalent-noise-bandwidth&catid=94:2014-06-june-articles&Itemid=189}
  {\bibfield  {journal} {\bibinfo  {journal} {High Freq. Electron. Mag.}\ ,\
  \bibinfo {pages} {1}} (\bibinfo {year} {2014})}\BibitemShut {NoStop}%
\bibitem [{\citenamefont {Romero}, \citenamefont {Garc\'{\i}a-Ripoll},\ and\
  \citenamefont {Solano}(2009)}]{PhysRevLett.102.173602}%
  \BibitemOpen
  \bibfield  {author} {\bibinfo {author} {\bibfnamefont {G.}~\bibnamefont
  {Romero}}, \bibinfo {author} {\bibfnamefont {J.~J.}\ \bibnamefont
  {Garc\'{\i}a-Ripoll}}, \ and\ \bibinfo {author} {\bibfnamefont
  {E.}~\bibnamefont {Solano}},\ }\href {\doibase
  10.1103/PhysRevLett.102.173602} {\bibfield  {journal} {\bibinfo  {journal}
  {Phys. Rev. Lett.}\ }\textbf {\bibinfo {volume} {102}},\ \bibinfo {pages}
  {173602} (\bibinfo {year} {2009})}\BibitemShut {NoStop}%
\bibitem [{\citenamefont {Inomata}\ \emph {et~al.}(2016)\citenamefont
  {Inomata}, \citenamefont {Lin}, \citenamefont {Koshino}, \citenamefont
  {Oliver}, \citenamefont {Tsai}, \citenamefont {Yamamoto},\ and\ \citenamefont
  {Nakamura}}]{inomata_single_2016}%
  \BibitemOpen
  \bibfield  {author} {\bibinfo {author} {\bibfnamefont {K.}~\bibnamefont
  {Inomata}}, \bibinfo {author} {\bibfnamefont {Z.}~\bibnamefont {Lin}},
  \bibinfo {author} {\bibfnamefont {K.}~\bibnamefont {Koshino}}, \bibinfo
  {author} {\bibfnamefont {W.~D.}\ \bibnamefont {Oliver}}, \bibinfo {author}
  {\bibfnamefont {J.-S.}\ \bibnamefont {Tsai}}, \bibinfo {author}
  {\bibfnamefont {T.}~\bibnamefont {Yamamoto}}, \ and\ \bibinfo {author}
  {\bibfnamefont {Y.}~\bibnamefont {Nakamura}},\ }\href {\doibase
  10.1038/ncomms12303} {\bibfield  {journal} {\bibinfo  {journal} {Nat.
  Commun.}\ }\textbf {\bibinfo {volume} {7}},\ \bibinfo {pages} {12303}
  (\bibinfo {year} {2016})}\BibitemShut {NoStop}%
\bibitem [{\citenamefont {Chen}\ \emph {et~al.}(2011)\citenamefont {Chen},
  \citenamefont {Hover}, \citenamefont {Sendelbach}, \citenamefont {Maurer},
  \citenamefont {Merkel}, \citenamefont {Pritchett}, \citenamefont {Wilhelm},\
  and\ \citenamefont {McDermott}}]{PhysRevLett.107.217401}%
  \BibitemOpen
  \bibfield  {author} {\bibinfo {author} {\bibfnamefont {Y.-F.}\ \bibnamefont
  {Chen}}, \bibinfo {author} {\bibfnamefont {D.}~\bibnamefont {Hover}},
  \bibinfo {author} {\bibfnamefont {S.}~\bibnamefont {Sendelbach}}, \bibinfo
  {author} {\bibfnamefont {L.}~\bibnamefont {Maurer}}, \bibinfo {author}
  {\bibfnamefont {S.~T.}\ \bibnamefont {Merkel}}, \bibinfo {author}
  {\bibfnamefont {E.~J.}\ \bibnamefont {Pritchett}}, \bibinfo {author}
  {\bibfnamefont {F.~K.}\ \bibnamefont {Wilhelm}}, \ and\ \bibinfo {author}
  {\bibfnamefont {R.}~\bibnamefont {McDermott}},\ }\href {\doibase
  10.1103/PhysRevLett.107.217401} {\bibfield  {journal} {\bibinfo  {journal}
  {Phys. Rev. Lett.}\ }\textbf {\bibinfo {volume} {107}},\ \bibinfo {pages}
  {217401} (\bibinfo {year} {2011})}\BibitemShut {NoStop}%
\bibitem [{\citenamefont {Poudel}, \citenamefont {McDermott},\ and\
  \citenamefont {Vavilov}(2012)}]{PhysRevB.86.174506}%
  \BibitemOpen
  \bibfield  {author} {\bibinfo {author} {\bibfnamefont {A.}~\bibnamefont
  {Poudel}}, \bibinfo {author} {\bibfnamefont {R.}~\bibnamefont {McDermott}}, \
  and\ \bibinfo {author} {\bibfnamefont {M.~G.}\ \bibnamefont {Vavilov}},\
  }\href {\doibase 10.1103/PhysRevB.86.174506} {\bibfield  {journal} {\bibinfo
  {journal} {Phys. Rev. B}\ }\textbf {\bibinfo {volume} {86}},\ \bibinfo
  {pages} {174506} (\bibinfo {year} {2012})}\BibitemShut {NoStop}%
\bibitem [{\citenamefont {Xu}\ \emph {et~al.}(2023)\citenamefont {Xu},
  \citenamefont {Zhou}, \citenamefont {Tan}, \citenamefont {Pan}, \citenamefont
  {Wen}, \citenamefont {Zhang}, \citenamefont {Zhou}, \citenamefont {Sun},
  \citenamefont {Chen}, \citenamefont {Zhou}, \citenamefont {Dai},
  \citenamefont {Chu},\ and\ \citenamefont {Hao}}]{D2NA00937D}%
  \BibitemOpen
  \bibfield  {author} {\bibinfo {author} {\bibfnamefont {Q.}~\bibnamefont
  {Xu}}, \bibinfo {author} {\bibfnamefont {Z.}~\bibnamefont {Zhou}}, \bibinfo
  {author} {\bibfnamefont {C.}~\bibnamefont {Tan}}, \bibinfo {author}
  {\bibfnamefont {X.}~\bibnamefont {Pan}}, \bibinfo {author} {\bibfnamefont
  {Z.}~\bibnamefont {Wen}}, \bibinfo {author} {\bibfnamefont {J.}~\bibnamefont
  {Zhang}}, \bibinfo {author} {\bibfnamefont {D.}~\bibnamefont {Zhou}},
  \bibinfo {author} {\bibfnamefont {Y.}~\bibnamefont {Sun}}, \bibinfo {author}
  {\bibfnamefont {X.}~\bibnamefont {Chen}}, \bibinfo {author} {\bibfnamefont
  {L.}~\bibnamefont {Zhou}}, \bibinfo {author} {\bibfnamefont {N.}~\bibnamefont
  {Dai}}, \bibinfo {author} {\bibfnamefont {J.}~\bibnamefont {Chu}}, \ and\
  \bibinfo {author} {\bibfnamefont {J.}~\bibnamefont {Hao}},\ }\href {\doibase
  10.1039/D2NA00937D} {\bibfield  {journal} {\bibinfo  {journal} {Nanoscale
  Adv.}\ }\textbf {\bibinfo {volume} {5}},\ \bibinfo {pages} {2054} (\bibinfo
  {year} {2023})}\BibitemShut {NoStop}%
\bibitem [{\citenamefont {Guarcello}\ \emph
  {et~al.}(2019{\natexlab{a}})\citenamefont {Guarcello}, \citenamefont
  {Braggio}, \citenamefont {Solinas}, \citenamefont {Pepe},\ and\ \citenamefont
  {Giazotto}}]{PhysRevApplied.11.054074}%
  \BibitemOpen
  \bibfield  {author} {\bibinfo {author} {\bibfnamefont {C.}~\bibnamefont
  {Guarcello}}, \bibinfo {author} {\bibfnamefont {A.}~\bibnamefont {Braggio}},
  \bibinfo {author} {\bibfnamefont {P.}~\bibnamefont {Solinas}}, \bibinfo
  {author} {\bibfnamefont {G.~P.}\ \bibnamefont {Pepe}}, \ and\ \bibinfo
  {author} {\bibfnamefont {F.}~\bibnamefont {Giazotto}},\ }\href {\doibase
  10.1103/PhysRevApplied.11.054074} {\bibfield  {journal} {\bibinfo  {journal}
  {Phys. Rev. Appl.}\ }\textbf {\bibinfo {volume} {11}},\ \bibinfo {pages}
  {054074} (\bibinfo {year} {2019}{\natexlab{a}})}\BibitemShut {NoStop}%
\bibitem [{\citenamefont {Gr\o{}nbech-Jensen}\ \emph
  {et~al.}(2004)\citenamefont {Gr\o{}nbech-Jensen}, \citenamefont {Castellano},
  \citenamefont {Chiarello}, \citenamefont {Cirillo}, \citenamefont {Cosmelli},
  \citenamefont {Filippenko}, \citenamefont {Russo},\ and\ \citenamefont
  {Torrioli}}]{PhysRevLett.93.107002}%
  \BibitemOpen
  \bibfield  {author} {\bibinfo {author} {\bibfnamefont {N.}~\bibnamefont
  {Gr\o{}nbech-Jensen}}, \bibinfo {author} {\bibfnamefont {M.~G.}\ \bibnamefont
  {Castellano}}, \bibinfo {author} {\bibfnamefont {F.}~\bibnamefont
  {Chiarello}}, \bibinfo {author} {\bibfnamefont {M.}~\bibnamefont {Cirillo}},
  \bibinfo {author} {\bibfnamefont {C.}~\bibnamefont {Cosmelli}}, \bibinfo
  {author} {\bibfnamefont {L.~V.}\ \bibnamefont {Filippenko}}, \bibinfo
  {author} {\bibfnamefont {R.}~\bibnamefont {Russo}}, \ and\ \bibinfo {author}
  {\bibfnamefont {G.}~\bibnamefont {Torrioli}},\ }\href {\doibase
  10.1103/PhysRevLett.93.107002} {\bibfield  {journal} {\bibinfo  {journal}
  {Phys. Rev. Lett.}\ }\textbf {\bibinfo {volume} {93}},\ \bibinfo {pages}
  {107002} (\bibinfo {year} {2004})}\BibitemShut {NoStop}%
\bibitem [{\citenamefont {Oelsner}\ \emph {et~al.}(2013)\citenamefont
  {Oelsner}, \citenamefont {Revin}, \citenamefont {Il'ichev}, \citenamefont
  {Pankratov}, \citenamefont {Meyer}, \citenamefont {Grönberg}, \citenamefont
  {Hassel},\ and\ \citenamefont {Kuzmin}}]{10.1063/1.4824308}%
  \BibitemOpen
  \bibfield  {author} {\bibinfo {author} {\bibfnamefont {G.}~\bibnamefont
  {Oelsner}}, \bibinfo {author} {\bibfnamefont {L.~S.}\ \bibnamefont {Revin}},
  \bibinfo {author} {\bibfnamefont {E.}~\bibnamefont {Il'ichev}}, \bibinfo
  {author} {\bibfnamefont {A.~L.}\ \bibnamefont {Pankratov}}, \bibinfo {author}
  {\bibfnamefont {H.-G.}\ \bibnamefont {Meyer}}, \bibinfo {author}
  {\bibfnamefont {L.}~\bibnamefont {Grönberg}}, \bibinfo {author}
  {\bibfnamefont {J.}~\bibnamefont {Hassel}}, \ and\ \bibinfo {author}
  {\bibfnamefont {L.~S.}\ \bibnamefont {Kuzmin}},\ }\href {\doibase
  10.1063/1.4824308} {\bibfield  {journal} {\bibinfo  {journal} {Appl. Phys.
  Lett.}\ }\textbf {\bibinfo {volume} {103}},\ \bibinfo {pages} {142605}
  (\bibinfo {year} {2013})}\BibitemShut {NoStop}%
\bibitem [{\citenamefont {Revin}\ \emph {et~al.}(2020)\citenamefont {Revin},
  \citenamefont {Pankratov}, \citenamefont {Gordeeva}, \citenamefont
  {Yablokov}, \citenamefont {Rakut}, \citenamefont {Zbrozhek},\ and\
  \citenamefont {Kuzmin}}]{REVIN2020960}%
  \BibitemOpen
  \bibfield  {author} {\bibinfo {author} {\bibfnamefont {L.~S.}\ \bibnamefont
  {Revin}}, \bibinfo {author} {\bibfnamefont {A.~L.}\ \bibnamefont
  {Pankratov}}, \bibinfo {author} {\bibfnamefont {A.~V.}\ \bibnamefont
  {Gordeeva}}, \bibinfo {author} {\bibfnamefont {A.~A.}\ \bibnamefont
  {Yablokov}}, \bibinfo {author} {\bibfnamefont {I.~V.}\ \bibnamefont {Rakut}},
  \bibinfo {author} {\bibfnamefont {V.~O.}\ \bibnamefont {Zbrozhek}}, \ and\
  \bibinfo {author} {\bibfnamefont {L.~S.}\ \bibnamefont {Kuzmin}},\ }\href
  {\doibase https://doi.org/10.3762/bjnano.11.80} {\bibfield  {journal}
  {\bibinfo  {journal} {Beilstein J. Nanotechnol.}\ }\textbf {\bibinfo {volume}
  {11}},\ \bibinfo {pages} {960} (\bibinfo {year} {2020})}\BibitemShut
  {NoStop}%
\bibitem [{\citenamefont {Pankratov}\ \emph
  {et~al.}(2022{\natexlab{b}})\citenamefont {Pankratov}, \citenamefont {Revin},
  \citenamefont {Gordeeva}, \citenamefont {Yablokov}, \citenamefont {Kuzmin},\
  and\ \citenamefont {Il’ichev}}]{pankratov_towards_2022}%
  \BibitemOpen
  \bibfield  {author} {\bibinfo {author} {\bibfnamefont {A.~L.}\ \bibnamefont
  {Pankratov}}, \bibinfo {author} {\bibfnamefont {L.~S.}\ \bibnamefont
  {Revin}}, \bibinfo {author} {\bibfnamefont {A.~V.}\ \bibnamefont {Gordeeva}},
  \bibinfo {author} {\bibfnamefont {A.~A.}\ \bibnamefont {Yablokov}}, \bibinfo
  {author} {\bibfnamefont {L.~S.}\ \bibnamefont {Kuzmin}}, \ and\ \bibinfo
  {author} {\bibfnamefont {E.}~\bibnamefont {Il’ichev}},\ }\href {\doibase
  10.1038/s41534-022-00569-5} {\bibfield  {journal} {\bibinfo  {journal} {Npj
  Quantum Inf.}\ }\textbf {\bibinfo {volume} {8}},\ \bibinfo {pages} {61}
  (\bibinfo {year} {2022}{\natexlab{b}})}\BibitemShut {NoStop}%
\bibitem [{\citenamefont {Guarcello}\ \emph
  {et~al.}(2019{\natexlab{b}})\citenamefont {Guarcello}, \citenamefont
  {Valenti}, \citenamefont {Spagnolo}, \citenamefont {Pierro},\ and\
  \citenamefont {Filatrella}}]{PhysRevApplied.11.044078}%
  \BibitemOpen
  \bibfield  {author} {\bibinfo {author} {\bibfnamefont {C.}~\bibnamefont
  {Guarcello}}, \bibinfo {author} {\bibfnamefont {D.}~\bibnamefont {Valenti}},
  \bibinfo {author} {\bibfnamefont {B.}~\bibnamefont {Spagnolo}}, \bibinfo
  {author} {\bibfnamefont {V.}~\bibnamefont {Pierro}}, \ and\ \bibinfo {author}
  {\bibfnamefont {G.}~\bibnamefont {Filatrella}},\ }\href {\doibase
  10.1103/PhysRevApplied.11.044078} {\bibfield  {journal} {\bibinfo  {journal}
  {Phys. Rev. Appl.}\ }\textbf {\bibinfo {volume} {11}},\ \bibinfo {pages}
  {044078} (\bibinfo {year} {2019}{\natexlab{b}})}\BibitemShut {NoStop}%
\bibitem [{\citenamefont {Ali}\ \emph {et~al.}(2024)\citenamefont {Ali},
  \citenamefont {Ouyang}, \citenamefont {He}, \citenamefont {Chai},\ and\
  \citenamefont {Wei}}]{ali_josephson_2024}%
  \BibitemOpen
  \bibfield  {author} {\bibinfo {author} {\bibfnamefont {S.}~\bibnamefont
  {Ali}}, \bibinfo {author} {\bibfnamefont {P.~H.}\ \bibnamefont {Ouyang}},
  \bibinfo {author} {\bibfnamefont {J.~X.}\ \bibnamefont {He}}, \bibinfo
  {author} {\bibfnamefont {Y.~Q.}\ \bibnamefont {Chai}}, \ and\ \bibinfo
  {author} {\bibfnamefont {L.~F.}\ \bibnamefont {Wei}},\ }\href {\doibase
  10.1038/s41598-024-52684-2} {\bibfield  {journal} {\bibinfo  {journal} {Sci.
  Rep.}\ }\textbf {\bibinfo {volume} {14}},\ \bibinfo {pages} {2531} (\bibinfo
  {year} {2024})}\BibitemShut {NoStop}%
\bibitem [{\citenamefont {Chai}\ \emph {et~al.}(2025)\citenamefont {Chai},
  \citenamefont {Wang}, \citenamefont {OuYang},\ and\ \citenamefont
  {Wei}}]{PhysRevB.111.024501}%
  \BibitemOpen
  \bibfield  {author} {\bibinfo {author} {\bibfnamefont {Y.~Q.}\ \bibnamefont
  {Chai}}, \bibinfo {author} {\bibfnamefont {S.~N.}\ \bibnamefont {Wang}},
  \bibinfo {author} {\bibfnamefont {P.~H.}\ \bibnamefont {OuYang}}, \ and\
  \bibinfo {author} {\bibfnamefont {L.~F.}\ \bibnamefont {Wei}},\ }\href
  {\doibase 10.1103/PhysRevB.111.024501} {\bibfield  {journal} {\bibinfo
  {journal} {Phys. Rev. B}\ }\textbf {\bibinfo {volume} {111}},\ \bibinfo
  {pages} {024501} (\bibinfo {year} {2025})}\BibitemShut {NoStop}%
\bibitem [{\citenamefont {Rettaroli}\ \emph {et~al.}(2021)\citenamefont
  {Rettaroli}, \citenamefont {Alesini}, \citenamefont {Babusci}, \citenamefont
  {Barone}, \citenamefont {Buonomo}, \citenamefont {Beretta}, \citenamefont
  {Castellano}, \citenamefont {Chiarello}, \citenamefont {Di~Gioacchino},
  \citenamefont {Felici}, \citenamefont {Filatrella}, \citenamefont {Foggetta},
  \citenamefont {Gallo}, \citenamefont {Gatti}, \citenamefont {Ligi},
  \citenamefont {Maccarrone}, \citenamefont {Mattioli}, \citenamefont {Pagano},
  \citenamefont {Tocci},\ and\ \citenamefont {Torrioli}}]{instruments5030025}%
  \BibitemOpen
  \bibfield  {author} {\bibinfo {author} {\bibfnamefont {A.}~\bibnamefont
  {Rettaroli}}, \bibinfo {author} {\bibfnamefont {D.}~\bibnamefont {Alesini}},
  \bibinfo {author} {\bibfnamefont {D.}~\bibnamefont {Babusci}}, \bibinfo
  {author} {\bibfnamefont {C.}~\bibnamefont {Barone}}, \bibinfo {author}
  {\bibfnamefont {B.}~\bibnamefont {Buonomo}}, \bibinfo {author} {\bibfnamefont
  {M.~M.}\ \bibnamefont {Beretta}}, \bibinfo {author} {\bibfnamefont
  {G.}~\bibnamefont {Castellano}}, \bibinfo {author} {\bibfnamefont
  {F.}~\bibnamefont {Chiarello}}, \bibinfo {author} {\bibfnamefont
  {D.}~\bibnamefont {Di~Gioacchino}}, \bibinfo {author} {\bibfnamefont
  {G.}~\bibnamefont {Felici}}, \bibinfo {author} {\bibfnamefont
  {G.}~\bibnamefont {Filatrella}}, \bibinfo {author} {\bibfnamefont {L.~G.}\
  \bibnamefont {Foggetta}}, \bibinfo {author} {\bibfnamefont {A.}~\bibnamefont
  {Gallo}}, \bibinfo {author} {\bibfnamefont {C.}~\bibnamefont {Gatti}},
  \bibinfo {author} {\bibfnamefont {C.}~\bibnamefont {Ligi}}, \bibinfo {author}
  {\bibfnamefont {G.}~\bibnamefont {Maccarrone}}, \bibinfo {author}
  {\bibfnamefont {F.}~\bibnamefont {Mattioli}}, \bibinfo {author}
  {\bibfnamefont {S.}~\bibnamefont {Pagano}}, \bibinfo {author} {\bibfnamefont
  {S.}~\bibnamefont {Tocci}}, \ and\ \bibinfo {author} {\bibfnamefont
  {G.}~\bibnamefont {Torrioli}},\ }\href {\doibase 10.3390/instruments5030025}
  {\bibfield  {journal} {\bibinfo  {journal} {Rev. Sci. Instrum.}\ }\textbf
  {\bibinfo {volume} {5}},\ \bibinfo {pages} {25} (\bibinfo {year}
  {2021})}\BibitemShut {NoStop}%
\bibitem [{\citenamefont {D'Elia}\ \emph {et~al.}(2023)\citenamefont {D'Elia},
  \citenamefont {Rettaroli}, \citenamefont {Tocci}, \citenamefont {Babusci},
  \citenamefont {Barone}, \citenamefont {Beretta}, \citenamefont {Buonomo},
  \citenamefont {Chiarello}, \citenamefont {Chikhi}, \citenamefont
  {Di~Gioacchino}, \citenamefont {Felici}, \citenamefont {Filatrella},
  \citenamefont {Fistul}, \citenamefont {Foggetta}, \citenamefont {Gatti},
  \citenamefont {Il'ichev}, \citenamefont {Ligi}, \citenamefont {Lisitskiy},
  \citenamefont {Maccarrone}, \citenamefont {Mattioli}, \citenamefont
  {Oelsner}, \citenamefont {Pagano}, \citenamefont {Piersanti}, \citenamefont
  {Ruggiero}, \citenamefont {Torrioli},\ and\ \citenamefont
  {Zagoskin}}]{DElia2023SteppingCT}%
  \BibitemOpen
  \bibfield  {author} {\bibinfo {author} {\bibfnamefont {A.}~\bibnamefont
  {D'Elia}}, \bibinfo {author} {\bibfnamefont {A.}~\bibnamefont {Rettaroli}},
  \bibinfo {author} {\bibfnamefont {S.}~\bibnamefont {Tocci}}, \bibinfo
  {author} {\bibfnamefont {D.}~\bibnamefont {Babusci}}, \bibinfo {author}
  {\bibfnamefont {C.}~\bibnamefont {Barone}}, \bibinfo {author} {\bibfnamefont
  {M.}~\bibnamefont {Beretta}}, \bibinfo {author} {\bibfnamefont
  {B.}~\bibnamefont {Buonomo}}, \bibinfo {author} {\bibfnamefont
  {F.}~\bibnamefont {Chiarello}}, \bibinfo {author} {\bibfnamefont
  {N.}~\bibnamefont {Chikhi}}, \bibinfo {author} {\bibfnamefont
  {D.}~\bibnamefont {Di~Gioacchino}}, \bibinfo {author} {\bibfnamefont
  {G.}~\bibnamefont {Felici}}, \bibinfo {author} {\bibfnamefont
  {G.}~\bibnamefont {Filatrella}}, \bibinfo {author} {\bibfnamefont
  {M.}~\bibnamefont {Fistul}}, \bibinfo {author} {\bibfnamefont
  {L.}~\bibnamefont {Foggetta}}, \bibinfo {author} {\bibfnamefont
  {C.}~\bibnamefont {Gatti}}, \bibinfo {author} {\bibfnamefont
  {E.}~\bibnamefont {Il'ichev}}, \bibinfo {author} {\bibfnamefont
  {C.}~\bibnamefont {Ligi}}, \bibinfo {author} {\bibfnamefont {M.}~\bibnamefont
  {Lisitskiy}}, \bibinfo {author} {\bibfnamefont {G.}~\bibnamefont
  {Maccarrone}}, \bibinfo {author} {\bibfnamefont {F.}~\bibnamefont
  {Mattioli}}, \bibinfo {author} {\bibfnamefont {G.}~\bibnamefont {Oelsner}},
  \bibinfo {author} {\bibfnamefont {S.}~\bibnamefont {Pagano}}, \bibinfo
  {author} {\bibfnamefont {L.}~\bibnamefont {Piersanti}}, \bibinfo {author}
  {\bibfnamefont {B.}~\bibnamefont {Ruggiero}}, \bibinfo {author}
  {\bibfnamefont {G.}~\bibnamefont {Torrioli}}, \ and\ \bibinfo {author}
  {\bibfnamefont {A.}~\bibnamefont {Zagoskin}},\ }\href {\doibase
  10.1109/TASC.2022.3218072} {\bibfield  {journal} {\bibinfo  {journal} {IEEE
  Trans. Appl. Supercond.}\ }\textbf {\bibinfo {volume} {33}},\ \bibinfo
  {pages} {1} (\bibinfo {year} {2023})}\BibitemShut {NoStop}%
\bibitem [{\citenamefont {Filatrella}\ \emph {et~al.}(2023)\citenamefont
  {Filatrella}, \citenamefont {Barone}, \citenamefont {Carapella},
  \citenamefont {Gatti}, \citenamefont {Granata}, \citenamefont {Guarcello},
  \citenamefont {Mauro}, \citenamefont {Komnang}, \citenamefont {Pierro},
  \citenamefont {Rettaroli},\ and\ \citenamefont {Pagano}}]{9919334}%
  \BibitemOpen
  \bibfield  {author} {\bibinfo {author} {\bibfnamefont {G.}~\bibnamefont
  {Filatrella}}, \bibinfo {author} {\bibfnamefont {C.}~\bibnamefont {Barone}},
  \bibinfo {author} {\bibfnamefont {G.}~\bibnamefont {Carapella}}, \bibinfo
  {author} {\bibfnamefont {C.}~\bibnamefont {Gatti}}, \bibinfo {author}
  {\bibfnamefont {V.}~\bibnamefont {Granata}}, \bibinfo {author} {\bibfnamefont
  {C.}~\bibnamefont {Guarcello}}, \bibinfo {author} {\bibfnamefont
  {C.}~\bibnamefont {Mauro}}, \bibinfo {author} {\bibfnamefont {A.~P.}\
  \bibnamefont {Komnang}}, \bibinfo {author} {\bibfnamefont {V.}~\bibnamefont
  {Pierro}}, \bibinfo {author} {\bibfnamefont {A.}~\bibnamefont {Rettaroli}}, \
  and\ \bibinfo {author} {\bibfnamefont {S.}~\bibnamefont {Pagano}},\ }\href
  {\doibase 10.1109/TASC.2022.3214500} {\bibfield  {journal} {\bibinfo
  {journal} {IEEE Trans. Appl. Supercond.}\ }\textbf {\bibinfo {volume} {33}},\
  \bibinfo {pages} {1} (\bibinfo {year} {2023})}\BibitemShut {NoStop}%
\bibitem [{\citenamefont {Cheng}\ \emph {et~al.}(2018)\citenamefont {Cheng},
  \citenamefont {Cirillo}, \citenamefont {Salina},\ and\ \citenamefont
  {Gr\o{}nbech-Jensen}}]{PhysRevE.98.012140}%
  \BibitemOpen
  \bibfield  {author} {\bibinfo {author} {\bibfnamefont {C.}~\bibnamefont
  {Cheng}}, \bibinfo {author} {\bibfnamefont {M.}~\bibnamefont {Cirillo}},
  \bibinfo {author} {\bibfnamefont {G.}~\bibnamefont {Salina}}, \ and\ \bibinfo
  {author} {\bibfnamefont {N.}~\bibnamefont {Gr\o{}nbech-Jensen}},\ }\href
  {\doibase 10.1103/PhysRevE.98.012140} {\bibfield  {journal} {\bibinfo
  {journal} {Phys. Rev. E}\ }\textbf {\bibinfo {volume} {98}},\ \bibinfo
  {pages} {012140} (\bibinfo {year} {2018})}\BibitemShut {NoStop}%
\bibitem [{\citenamefont {Guarcello}\ \emph {et~al.}(2021)\citenamefont
  {Guarcello}, \citenamefont {Piedjou~Komnang}, \citenamefont {Barone},
  \citenamefont {Rettaroli}, \citenamefont {Gatti}, \citenamefont {Pagano},\
  and\ \citenamefont {Filatrella}}]{PhysRevApplied.16.054015}%
  \BibitemOpen
  \bibfield  {author} {\bibinfo {author} {\bibfnamefont {C.}~\bibnamefont
  {Guarcello}}, \bibinfo {author} {\bibfnamefont {A.~S.}\ \bibnamefont
  {Piedjou~Komnang}}, \bibinfo {author} {\bibfnamefont {C.}~\bibnamefont
  {Barone}}, \bibinfo {author} {\bibfnamefont {A.}~\bibnamefont {Rettaroli}},
  \bibinfo {author} {\bibfnamefont {C.}~\bibnamefont {Gatti}}, \bibinfo
  {author} {\bibfnamefont {S.}~\bibnamefont {Pagano}}, \ and\ \bibinfo {author}
  {\bibfnamefont {G.}~\bibnamefont {Filatrella}},\ }\href {\doibase
  10.1103/PhysRevApplied.16.054015} {\bibfield  {journal} {\bibinfo  {journal}
  {Phys. Rev. Appl.}\ }\textbf {\bibinfo {volume} {16}},\ \bibinfo {pages}
  {054015} (\bibinfo {year} {2021})}\BibitemShut {NoStop}%
\bibitem [{\citenamefont {Piedjou~Komnang}\ \emph {et~al.}(2021)\citenamefont
  {Piedjou~Komnang}, \citenamefont {Guarcello}, \citenamefont {Barone},
  \citenamefont {Pagano},\ and\ \citenamefont {Filatrella}}]{9555447}%
  \BibitemOpen
  \bibfield  {author} {\bibinfo {author} {\bibfnamefont {A.~S.}\ \bibnamefont
  {Piedjou~Komnang}}, \bibinfo {author} {\bibfnamefont {C.}~\bibnamefont
  {Guarcello}}, \bibinfo {author} {\bibfnamefont {C.}~\bibnamefont {Barone}},
  \bibinfo {author} {\bibfnamefont {S.}~\bibnamefont {Pagano}}, \ and\ \bibinfo
  {author} {\bibfnamefont {G.}~\bibnamefont {Filatrella}},\ }in\ \href@noop {}
  {\emph {\bibinfo {booktitle} {2021 IEEE 14th Workshop on Low Temperature
  Electronics (WOLTE)}}}\ (\bibinfo {year} {2021})\ pp.\ \bibinfo {pages}
  {1--4}\BibitemShut {NoStop}%
\bibitem [{\citenamefont {Cui}\ \emph {et~al.}(2008)\citenamefont {Cui},
  \citenamefont {Lin}, \citenamefont {Yu}, \citenamefont {Peng}, \citenamefont
  {Zhu}, \citenamefont {Jing}, \citenamefont {Zheng}, \citenamefont {Lu},\ and\
  \citenamefont {Zhao}}]{2008_Cui}%
  \BibitemOpen
  \bibfield  {author} {\bibinfo {author} {\bibfnamefont {D.}~\bibnamefont
  {Cui}}, \bibinfo {author} {\bibfnamefont {D.}~\bibnamefont {Lin}}, \bibinfo
  {author} {\bibfnamefont {H.}~\bibnamefont {Yu}}, \bibinfo {author}
  {\bibfnamefont {Z.}~\bibnamefont {Peng}}, \bibinfo {author} {\bibfnamefont
  {X.}~\bibnamefont {Zhu}}, \bibinfo {author} {\bibfnamefont {X.}~\bibnamefont
  {Jing}}, \bibinfo {author} {\bibfnamefont {D.}~\bibnamefont {Zheng}},
  \bibinfo {author} {\bibfnamefont {L.}~\bibnamefont {Lu}}, \ and\ \bibinfo
  {author} {\bibfnamefont {S.}~\bibnamefont {Zhao}},\ }\href@noop {} {\bibfield
   {journal} {\bibinfo  {journal} {Acta Phys. Sin.}\ }\textbf {\bibinfo
  {volume} {57}},\ \bibinfo {pages} {5933} (\bibinfo {year}
  {2008})}\BibitemShut {NoStop}%
\bibitem [{\citenamefont {B\"uttiker}, \citenamefont {Harris},\ and\
  \citenamefont {Landauer}(1983)}]{PhysRevB.28.1268}%
  \BibitemOpen
  \bibfield  {author} {\bibinfo {author} {\bibfnamefont {M.}~\bibnamefont
  {B\"uttiker}}, \bibinfo {author} {\bibfnamefont {E.~P.}\ \bibnamefont
  {Harris}}, \ and\ \bibinfo {author} {\bibfnamefont {R.}~\bibnamefont
  {Landauer}},\ }\href {\doibase 10.1103/PhysRevB.28.1268} {\bibfield
  {journal} {\bibinfo  {journal} {Phys. Rev. B}\ }\textbf {\bibinfo {volume}
  {28}},\ \bibinfo {pages} {1268} (\bibinfo {year} {1983})}\BibitemShut
  {NoStop}%
\bibitem [{\citenamefont {Caldeira}\ and\ \citenamefont
  {Leggett}(1981)}]{PhysRevLett.46.211}%
  \BibitemOpen
  \bibfield  {author} {\bibinfo {author} {\bibfnamefont {A.~O.}\ \bibnamefont
  {Caldeira}}\ and\ \bibinfo {author} {\bibfnamefont {A.~J.}\ \bibnamefont
  {Leggett}},\ }\href {\doibase 10.1103/PhysRevLett.46.211} {\bibfield
  {journal} {\bibinfo  {journal} {Phys. Rev. Lett.}\ }\textbf {\bibinfo
  {volume} {46}},\ \bibinfo {pages} {211} (\bibinfo {year} {1981})}\BibitemShut
  {NoStop}%
\bibitem [{\citenamefont {Han}\ \emph {et~al.}(2021)\citenamefont {Han},
  \citenamefont {Ouyang}, \citenamefont {Li}, \citenamefont {Wang},\ and\
  \citenamefont {Wei}}]{han_experimentally_2021}%
  \BibitemOpen
  \bibfield  {author} {\bibinfo {author} {\bibfnamefont {J.~G.}\ \bibnamefont
  {Han}}, \bibinfo {author} {\bibfnamefont {P.~H.}\ \bibnamefont {Ouyang}},
  \bibinfo {author} {\bibfnamefont {E.~P.}\ \bibnamefont {Li}}, \bibinfo
  {author} {\bibfnamefont {Y.~W.}\ \bibnamefont {Wang}}, \ and\ \bibinfo
  {author} {\bibfnamefont {L.~F.}\ \bibnamefont {Wei}},\ }\href {\doibase
  10.7498/aps.70.20210393} {\bibfield  {journal} {\bibinfo  {journal} {Acta
  Phys. Sin.}\ }\textbf {\bibinfo {volume} {70}},\ \bibinfo {pages} {170304}
  (\bibinfo {year} {2021})}\BibitemShut {NoStop}%
\bibitem [{\citenamefont {Ouyang}\ \emph {et~al.}(2024)\citenamefont {Ouyang},
  \citenamefont {He}, \citenamefont {Wang}, \citenamefont {Chai}, \citenamefont
  {He}, \citenamefont {Chang},\ and\ \citenamefont
  {Wei}}]{PhysRevResearch.6.013236}%
  \BibitemOpen
  \bibfield  {author} {\bibinfo {author} {\bibfnamefont {P.~H.}\ \bibnamefont
  {Ouyang}}, \bibinfo {author} {\bibfnamefont {S.~R.}\ \bibnamefont {He}},
  \bibinfo {author} {\bibfnamefont {Y.~Z.}\ \bibnamefont {Wang}}, \bibinfo
  {author} {\bibfnamefont {Y.~Q.}\ \bibnamefont {Chai}}, \bibinfo {author}
  {\bibfnamefont {J.~X.}\ \bibnamefont {He}}, \bibinfo {author} {\bibfnamefont
  {H.}~\bibnamefont {Chang}}, \ and\ \bibinfo {author} {\bibfnamefont {L.~F.}\
  \bibnamefont {Wei}},\ }\href {\doibase 10.1103/PhysRevResearch.6.013236}
  {\bibfield  {journal} {\bibinfo  {journal} {Phys. Rev. Res.}\ }\textbf
  {\bibinfo {volume} {6}},\ \bibinfo {pages} {013236} (\bibinfo {year}
  {2024})}\BibitemShut {NoStop}%
\end{thebibliography}%

\appendix

\section{Device fabrication}
\begin{figure}[htbp]
  \centering
  \includegraphics[width=0.9\linewidth]{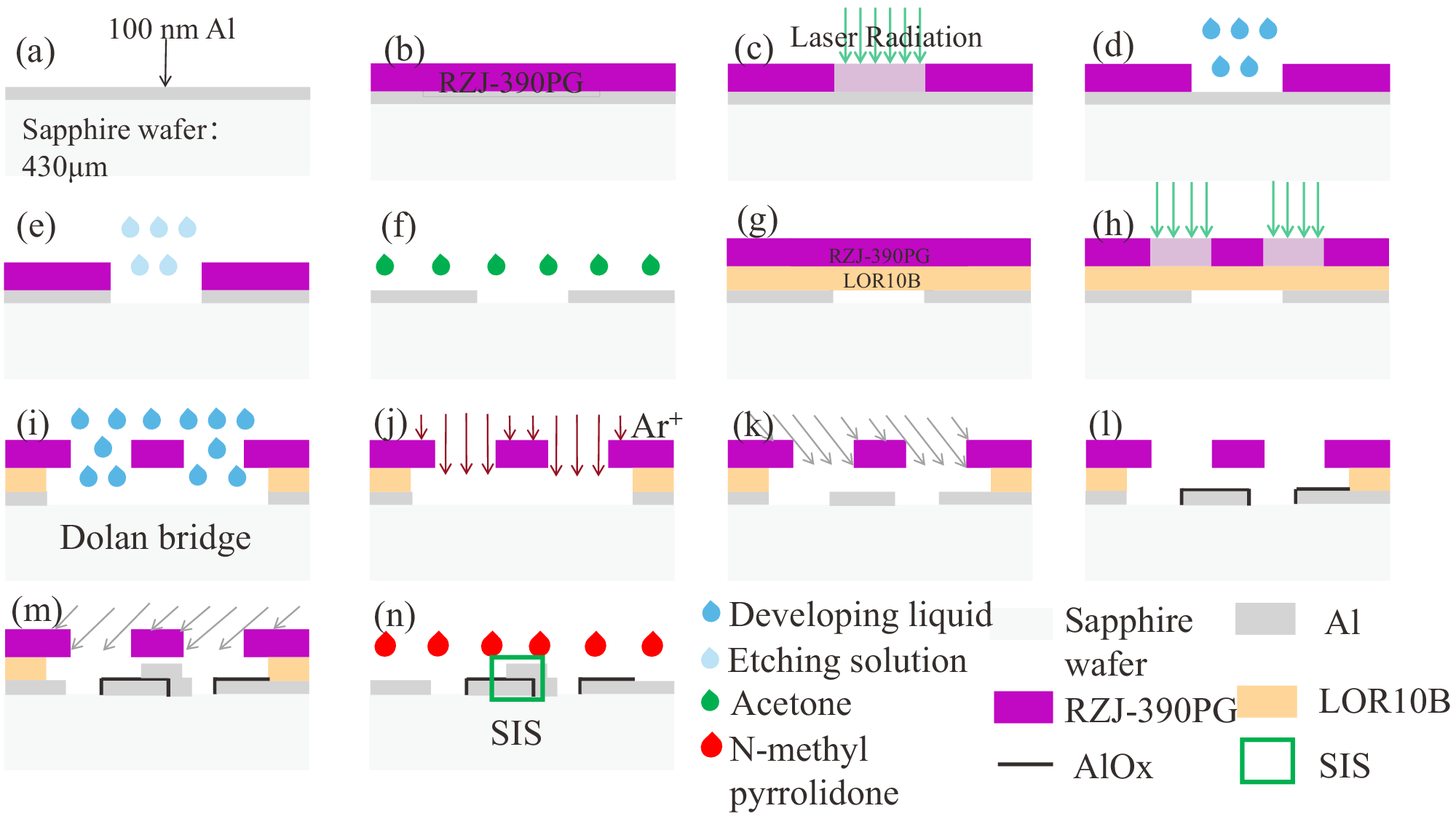}
  \caption{Preparation process of Al/AlOx/Al Josephson junctions, see the text for the explanation, in details.}
  \label{fig:JJ_make}
\end{figure}

The fabrication of a JJ is simply shown in Fig.~\ref{fig:JJ_make}. 
First, a 2-inch sapphire wafer with a $\langle0001\rangle$ crystal orientation and a thickness of 430~$\mu$m was cleaned by using acetone, isopropanol, and was baked at $180~\rm ^\circ C$ for 2 hours (to remove surface impurities and oxides for enhancing the adhesion of the Al film) and then was placed into the electron beam evaporation chamber. As shown in (a) of Fig.~\ref{fig:JJ_make}, a 100~nm aluminum film was then evaporated on the sapphire substrate in a high-vacuum environment ($\sim 10^{-5}$~Pa) with the coating rate being set as 0.3~nm/s.

Then, the preparation of the underlying circuitry was completed after the spin coating (b), photolithography (c), development (d), wet etching (e), and acetone-free removal (f), shown sequentially in Fig.~7. Then, the “Dolan” bridge structure was prepared by using the spin coating (g), photolithography (h), and development (i). This is an essential step before the later preparation of the Al/AlOx/Al Josephson junction structures using the oblique angle electron beam evaporation technique. Next, the sapphire substrate that has completed the above steps was put into an electron beam evaporation coater again, and then the plasma dry etching to remove the left impurities such as oxides. The Al/AlOx/Al junctions were prepared by using oblique-angle $54^\circ$ coating (100~nm), oxidizing (2.6~Pa,10~min) and $-54^\circ$ (120~nm) coatings, which corresponds to (j)$\sim$(m) in Fig.~\ref{fig:JJ_make}. Finally, removing glue with acetone and N-Methyl-2-Pyrrolidone was performed to obtain the SIS junction shown specifically in Fig.~\ref{fig:JJ_make}(n). An overall schematic as well as a physical drawing of the prepared Josephson junction is shown in Fig.~3. The width of the transmission line is about 10~$\mu$m, and the area of the Josephson junction is 0.78~$\mu m^{2}$. The microwave transmission line is directly coupled to the Josephson junction, and the microwave signals flow from the upper and lower ports through the transmission line and the Josephson junction in the middle. The distance between the microwave transmission line and the grounding line is set as 5~$\mu$m for the impedance matching with the external microwave signals $Z=50~\Omega$.

\section{Experimental cryogenic measurement circuit}
The Josephson junction devices packaged tested at an ambient temperature of 20~mK, provided by a Leiden Cryogenics B.V. dilution chiller (model CRYOGEN-FREE DILUTION REFRIGERATOR, CF-CS50). The samples were placed in a superconducting aluminum box, which effectively suppresses RF interference at cryogenic temperatures, and mounted onto a 20~mK disk of the dilution chiller.

The measurement results of the IV characteristic curve of the junction are shown in Fig.~4(a), based on the measurement setup in Fig.~8. The voltage signal $V_b$ generated by the Agilent 33250A arbitrary waveform generator is converted into a current signal through an adjustable resistor $R_L$ (ranging from 1~k$\Omega$ to 1~M$\Omega$), which serves as the external bias for the junction. 
The signal is displayed on the Agilent DSO71044 oscilloscope for visualization, and the data is transferred to a PC via an A/D acquisition card. The adjustable resistor $R_L$ is typically much larger than the resistance of the Josephson junction $R_J$, allowing the bias current to be calculated as $I_b = V_b/R_L$, with the junction's resistance considered negligible. The voltage signals at both ends of the junction are amplified by a room-temperature amplifier (500x gain) and measured using the Agilent DSO71044 oscilloscope. The data is then transferred to a PC via an A/D acquisition card.
\begin{figure}[htbp]
  \centering
   \includegraphics[width=8.3cm]{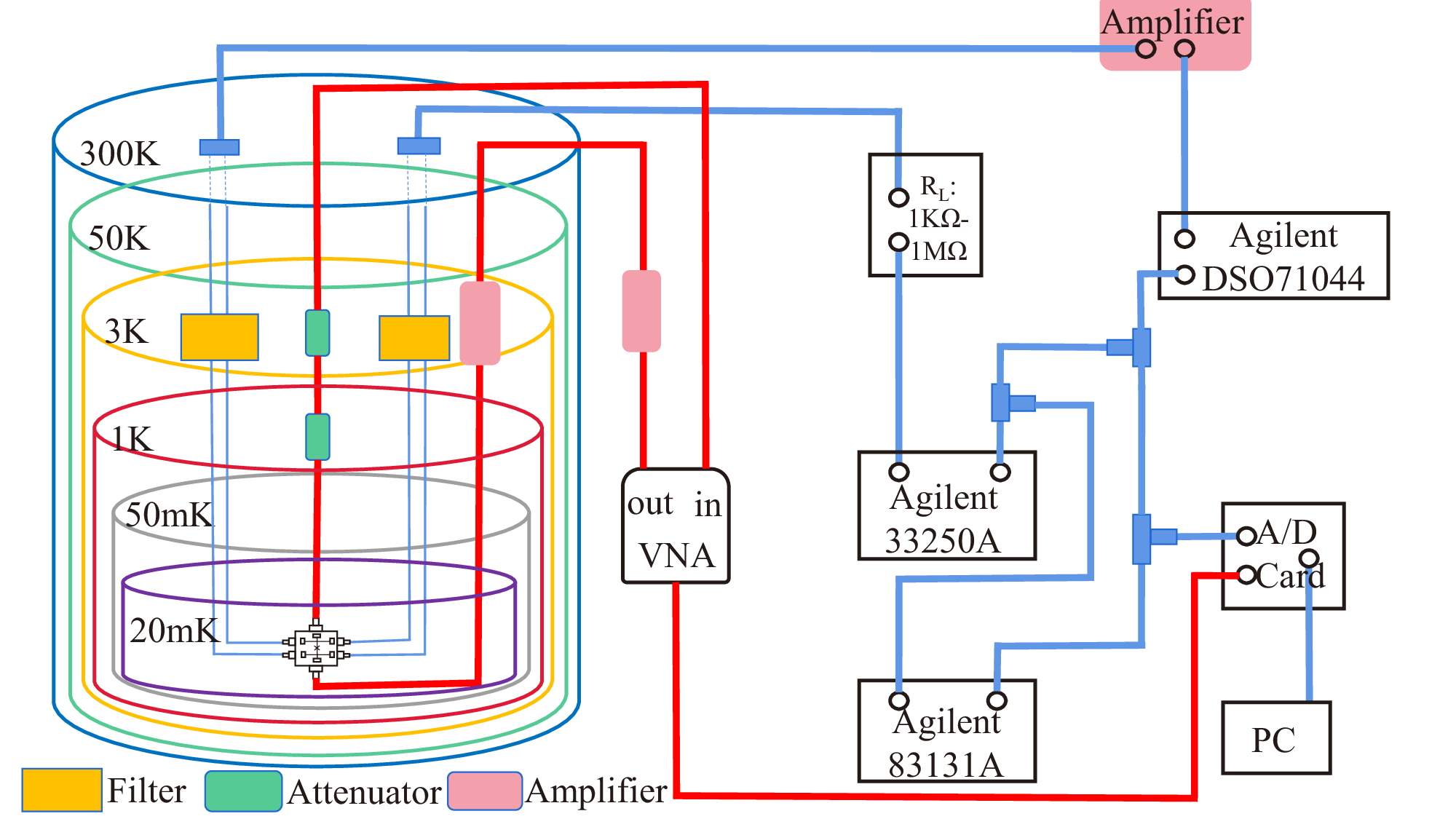}
  \caption{Experimental measurement setup. The dilution chiller creates a low-temperature environment of 50 mK, with the encapsulated sample placed on a 20~mK disk. The actual temperature of the sample is 50~mK due to the age of the chiller. The blue four-terminal line is used for standard four-terminal measurements, while the yellow section represents the filter. The red line carries the microwave signal input and output from the VNA. The VNA input line at the 3~K disk passes through an attenuator (green) with an attenuation of -10~dBm, while the 1 K disk has an attenuation of -37 to -40~dBm. The total attenuation on the input side is approximately -50~dBm, while the output side has a low-temperature amplifier with a gain of 30~dBm, and an ambient amplifier with a gain of 20~dBm, resulting in a total amplification of 50~dBm.}
  \label{fig:JJ_measure}
\end{figure}
The switched current measurement method is essentially the same as that used for the IV curve. The difference is that the bias current, provided by the arbitrary waveform generator (Agilent 33250A) and adjustable resistor, is a triangular wave. The triangular wave signal increases from zero to its peak value (slightly larger than the junction's critical current) at a constant rate of $dI/dt=56.5\times10^{-6}$~A/s. A counter (Agilent 83131A) measures the time at which the voltage signals at the ends of the junction appear, indicating the switched duration. The bias current is then rapidly reduced to zero, and a half cycle has waited to bring the junction back to the superconducting state from the non-zero voltage state. This wait minimizes interference with the switched duration measurement when the next triangular wave is driven. The synchronization (sync) signal from the arbitrary waveform generator (Agilent 33250A) is connected to a counter (Agilent 83131A), as shown in Fig.~8, to ensure accurate switched duration measurement.

The microwave signal measurement circuit involves the connection of the vector network analyzer (VNA), shown by the red lines in Fig.~8. In order to perform the $S_{21}$-parameter sweeps and provide microwave signals with different levels of attenuation required for experiments, the microwave signal from the VNA passes through attenuators on the 3~K and 1~K stages before reaching the junction's input. It interacts with the junction, passes through the output, and is amplified by both a cryogenic amplifier and an ambient amplifier before returning to the VNA.

\end{document}